\documentclass[graybox, envcountchap]{svmult}

\usepackage{mathptmx}        
\usepackage{amsmath}
\usepackage{amssymb}
\usepackage[table]{xcolor}
\usepackage{helvet}          
\usepackage{courier}         
\usepackage{dirtree}

\usepackage{makeidx}        
\usepackage{graphicx}        
\usepackage{subfig}

\usepackage{multicol}        
\usepackage[bottom]{footmisc}

\usepackage[]{natbib} 

\usepackage[breaklinks=true,colorlinks=true,linkcolor=blue,citecolor=blue,urlcolor=blue]{hyperref} 
\hypersetup{colorlinks=true,urlcolor=blue}
\providecommand{\Letter}{\textsuperscript{*}}
\makeindex             

\newcommand{\ml}{$ML$}
\newcommand{\infer}{$\mathcal{I}$}
\newcommand{\proppredict}{$\mathcal{P}_{t \to t+\Delta t}$}
\newcommand{\propcouple}{$\mathcal{P}_{1\to 2}$}
\newcommand{\emulate}{$\mathcal{E}$}
\newcommand{\control}{$\mathcal{C}$}

\newcommand\ion[2]{#1$\,${\scshape{#2}}}

\definecolor{badgeblue}{rgb}{0.1, 0.3, 0.6}
\definecolor{badgegreen}{rgb}{0.1, 0.5, 0.2}
\definecolor{badgeorange}{rgb}{0.8, 0.4, 0.0}

\begin{document}


\title{Machine Learning Applications in the Solar Interior and Atmosphere}
\author{Vishal Upendran and Carlos J. Díaz Baso}
\institute{Vishal Upendran (\Letter) \at SETI Institute, Mountain View, CA, USA 94043; Lockheed Martin Solar and Astrophysics Laboratory, Palo Alto, CA, USA 94304, \email{vupendran@seti.org}
\and Carlos J. Díaz Baso \at Institute of Theoretical Astrophysics, University of Oslo, P.O. Box 1029 Blindern, N-0315 Oslo, Norway; Rosseland Centre for Solar Physics, University of Oslo, P.O. Box 1029 Blindern, N-0315 Oslo, Norway \email{c.j.d.baso@astro.uio.no}}
%
%
\maketitle

\abstract{The solar interior and atmosphere constitute a strongly coupled, multiscale dynamical system defined by continuous temporal evolution. Predicting the state of this system requires modeling physical interactions that span disparate spatial and temporal scales, constrained by heterogeneous and sparse data. This chapter formalizes the Sun within the mathematical framework of dynamical systems theory, defining a global state vector ($\vec{S}$) governed by thermodynamical, kinetic, and magnetic properties. Because solar observations represent projected, multi-modal, and partially observable measurements ($\vec{O}[\vec{S}]$) of this hidden state, pure physics-based deduction is fundamentally constrained. We present a unifying thesis: the primary role of machine learning in heliophysics is to act as a structured family of mathematical operators---specifically Inference ($\mathcal{I}$), Emulation ($\mathcal{E}$), Propagation ($\mathcal{P}$), and Control ($\mathcal{C}$)---that reconstruct, simulate, and couple this global state vector across disparate physical layers. By framing the vast, multi-disciplinary literature within this unified state-space taxonomy, we demonstrate how data-driven techniques bridge observational gaps to resolve hidden multiscale couplings and dramatically accelerate the simulation of non-linear processes, establishing a rigorous pathway to predict the global evolution of the Sun.}

\section{The Sun as a multiscale dynamical system}
\label{sec:intro}

One of the nearest astrophysical objects of study available to mankind is our very own Sun. It is a star of spectral type G2V, with a radius of about $6.9 \times 10^8$ meters and a mass of about $2 \times 10^{30}$ kilograms, located at a distance of about $1.5 \times 10^{11}$ meters from Earth as the source of almost all the energy that sustains life. It is a complex system with several layers, each with its own unique properties. Several phenomena occur all across the solar interior and atmosphere, and depending on the time scale of each phenomenon, they may be coupled to the global evolution of the Sun. 

While we have a general understanding of the different layers of the Sun, the physical processes occurring in each layer are deeply intertwined. The solar interior consists of the core, the radiative and convective zones. Nuclear fusion in the core generates energy, which is then transported outwards through the other zones. Photons, our main source of information about the Sun, are generated in the core and then undergo a random walk through the radiative zone, taking about $10^5$ years to reach the convective zone. In the convective zone, plasma motion becomes dominant, and energy is transported by convection. Depending on the density stratification, the photons become free streaming, and are able to escape the Sun, giving us a direct view of the solar surface, which we call the photosphere. The photosphere is the visible surface of the Sun, and it is where most of the solar radiation is emitted. Above the photosphere, the plasma density decreases, and we observe different layers of the solar atmosphere. Further higher up, the plasma becomes free streaming, and we observe the solar wind, which is a stream of charged particles that flows outwards from the Sun and fills the heliosphere. 

\begin{figure}
    \centering
    \includegraphics[width=\linewidth]{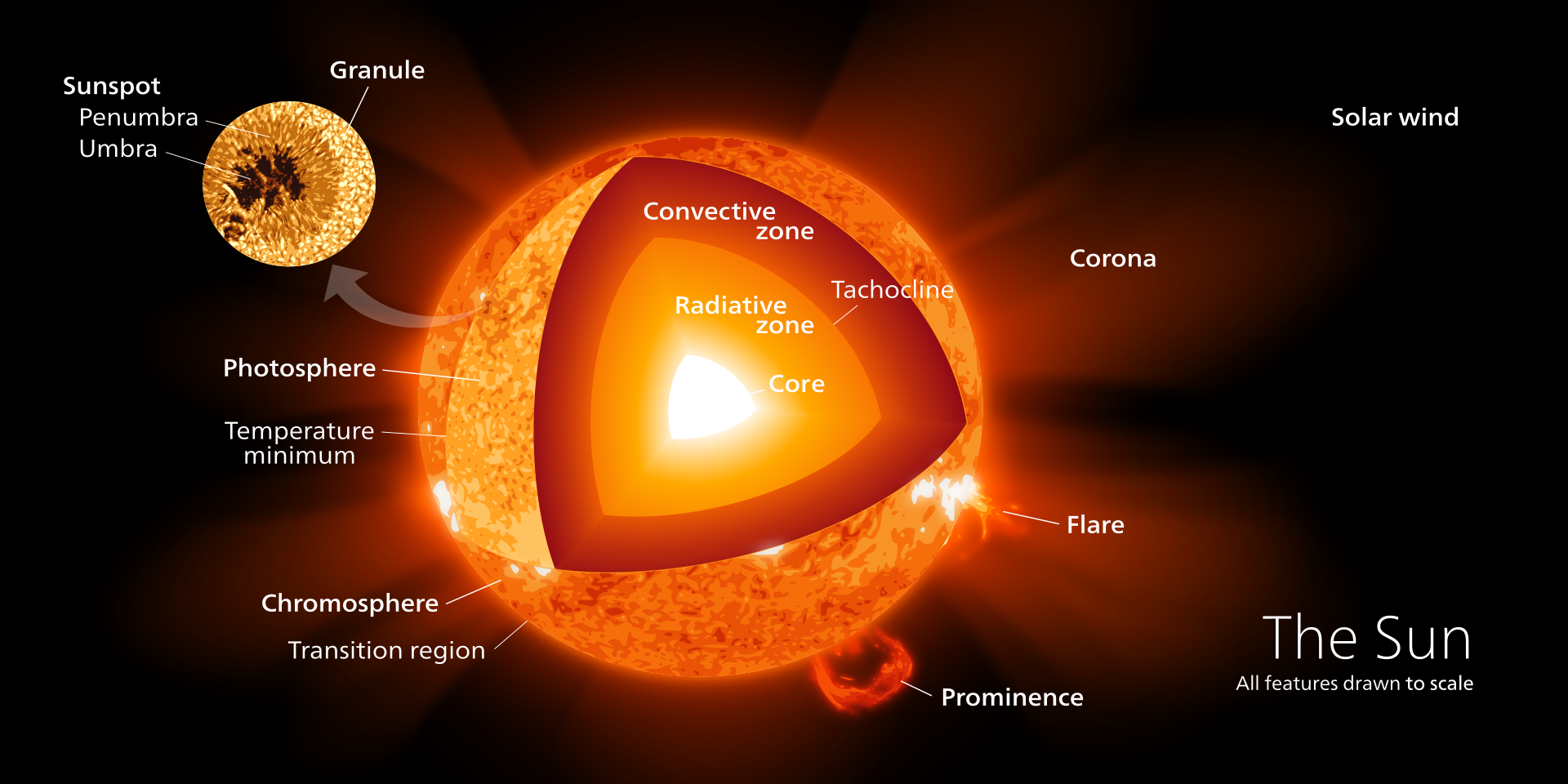}
    \caption{An illustration of different layers of the Sun. Art by Kelvinsong - Own work, CC BY-SA 3.0. Taken from Wikimedia commons.}
    \label{fig:sunlayer}
\end{figure}

The emission from the Sun is mostly a blackbody spectrum, but it exhibits a wide range of spectral lines corresponding to the transitions of ionization states of different elements. Each of these elements, and their ionization fractions depend on the specific layer of the solar atmosphere, and thus serve as diagnostics of the physical conditions in that layer. The solar atmosphere starts off from the photosphere, where the different elements are not quite fully ionized. With height, the plasma shows a mix of ions and neutrals in the chromosphere, transition region, and becomes almost fully ionized in the solar corona. Spectral observations of the different layers of the solar atmosphere suggest that the temperature in the solar atmosphere increases with height, reaching more than $10^6$ K in the solar corona, which is a long-standing puzzle in solar physics called the~\emph{Coronal Heating problem}. However, this is not a simple local phenomenon -- a complete understanding of the solar atmosphere requires understanding the coupling between the different layers. This coupling includes the solar interior setting the boundary conditions for the atmosphere, as well as the dynamic interactions between the atmospheric layers themselves\cite{priest}. A general graphic depicting the layers of the Sun is shown in Fig.~\ref{fig:sunlayer}.

The interior of the Sun is not directly observable through photons. While neutrino observations are potentially able to probe it, the more standard way of estimating interior properties is through helioseismology, which studies the propagation of acoustic waves. The acoustic waves are generated by turbulent convection in the convective zone, and they propagate through the solar interior, being refracted by the density stratification. By measuring the travel times of these waves at the solar surface, we can infer the properties of the solar interior, such as the sound speed, density, and flow velocities~\cite{Charbonneau}. 
\begin{figure}[ht!]
    \centering
    \includegraphics[width=\linewidth]{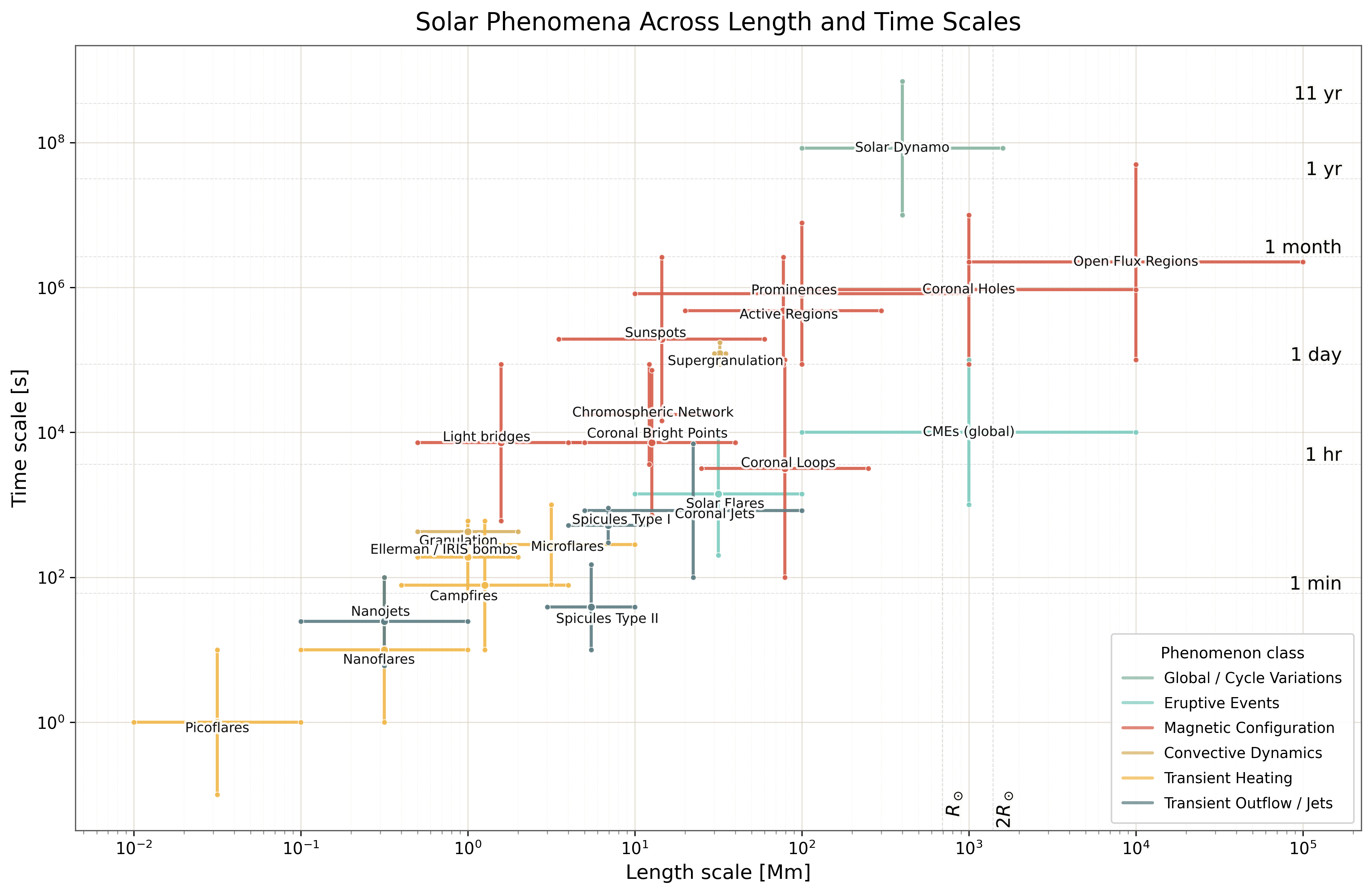}
    \caption[Length and time scales of solar phenomena]{An illustration of the length and time scales of various phenomena observed in the Sun. The plotted ranges are compiled from the phenomenon--reference associations in the source table: granulation and supergranulation~\cite{rincon}; chromospheric network~\cite{priest}; sunspots~\cite{solanki}; light bridges~\cite{lagg}; active regions~\cite{canfield}; Type I and Type II spicules~\cite{Zaqarashvili}; nanojets, picoflares, and nanoflares~\cite{Klimchuk,Parker}; coronal jets~\cite{Raouafi}; campfires~\cite{Berghmans}; Ellerman bombs and IRIS bombs~\cite{Ellerman,Peter}; microflares~\cite{Hannah}; coronal bright points~\cite{Madjarska}; coronal loops~\cite{Reale}; solar flares~\cite{Shibata}; CME source regions and global CMEs~\cite{Chen}; prominences~\cite{Parenti}; coronal holes and open flux regions~\cite{Cranmer}; and the solar dynamo~\cite{Charbonneau}.}
    \label{fig:scales}
\end{figure}

Plasma motion from the interior also brings magnetic fields to the surface. The different structures seen in the solar atmosphere are strongly modulated by the magnetic field, which is thought to be generated internally through a dynamo process. The interaction between plasma, magnetic field, and radiation gives rise to the morphological structure and phenomenon observed in the solar atmosphere. These phenomena manifest across length, time, and energy scales spanning about 10 orders of magnitude. Crucially, they can be coupled across all of these different scales. Hence, efforts seek to understand the Sun in a multiscale manner, balancing detailed studies of specific phenomena with the global evolution of the star. Fig.~\ref{fig:scales} displays the set of different known phenomena in the solar interior and atmosphere, and their characteristic length and time scales. Each color scale corresponds to the \emph{kind} of phenomena for which we mark the length and time scales. From Fig. ~\ref{fig:scales}, we see that the Sun exhibits phenomena ranging across $\sim$7 orders of magnitude in spatial extent, and $\sim$8 orders of magnitude in time. Not depicted in this figure is the range of energy scales measured, which can be upto $\sim$9 orders of magnitude~\cite{2004psci.book.....A}.

\begin{svgraybox}
This multiscale nature motivates us to think of the Sun as a \emph{system of systems}, where each subsystem is described by physical variables such as \emph{density, velocity, magnetic field}, and thermodynamic state. We collect these variables into a \emph{state vector} ($\vec{S}$), whose components are inferred from observations and serve as the foundation for our analytical operators (\infer, \proppredict, \propcouple, \emulate, \control). The coupling between subsystem state vectors gives rise to the phenomena observed across the solar interior and atmosphere.
\end{svgraybox}

The proximity of the Sun to Earth allows us to study it in much greater detail than any other star. Inherently, it is an excellent laboratory to study not just solar phenomena, but general astrophysical plasmas at great detail. As a consequence, we have a wealth of data available to study the Sun, ranging from remote sensing observations of the solar atmosphere and the heliosphere across the electromagnetic spectrum to in-situ measurements of the solar wind and energetic particles. This results in a very rich \emph{multi-messenger} data set that allows us to probe the Sun at different scales, presenting strong observational constraints on the physical models, and hence allowing us to test unsolved problems at very high precision.

With the ever growing data volume in solar physics, it is becoming increasingly difficult to analyze and interpret the data using traditional methods. Thus, the field of solar physics is increasingly turning towards machine learning ({\ml}) to analyze and interpret the data~\cite{camporeale2018machine,monica_bobra_2021_4521380,asensio2023machine, 2025PhyU...68..374I}.

This chapter focuses on the remote-sensing of source phenomena in the solar interior and atmosphere, while the subsequent chapter addresses the heliosphere and \emph{in-situ} processes.

\section{State vectors and {\ml} operators}
\label{sec:taxonomy}

The state-vector formalism gives a common language for organizing {\ml} applications across the solar interior and atmosphere. {\it The central object is not the algorithm, but the physical state on which the algorithm acts.} A convolutional network, transformer, neural operator, diffusion model, or random forest has a different scientific meaning depending on its target. Its physical relevance changes whether it estimates subsurface flow, advances a magnetogram, reconstructs a coronal volume, or forecasts an eruptive event. We therefore organize the literature in a hierarchy whose first level is the state vector itself and whose second level is the {\ml} operation applied to that state vector, defined below.

Let $\vec{S}$ denote the physical state vector of a solar subsystem. Typical variables may include \emph{Magneto hydrodynamic (MHD) variables}, like density, velocity, temperature, magnetic field, or particle / species population in different levels, or even parameters like line-profile parameters, morphology, or reduced event variables. 

Each state vector gives rise to an observable vector $\vec{O}$ (e.g., travel times, spectropolarimetric profiles, or intensity maps) measured by instruments, which is related to the physical state through a forward observation operator $\mathcal{O}$ such that $\vec{O} = \mathcal{O}[\vec{S}]$. Telescopes do not generally measure the precise state vector directly; instead, they measure projections, line-of-sight integrals, spectra, polarization states, image intensities, or helioseismic travel times. Hence, typically, the forward observation operator may be broken up into a combination of an observable synthesis ($\mathcal{O}_{\mathrm{phys}}$) and the telescope observation operator ($\mathcal{O}_{\mathrm{instr}}$), as $\mathcal{O} = \mathcal{O}_{\mathrm{instr}}[\mathcal{O}_{\mathrm{phys}}[\cdot]]$. 
\begin{svgraybox}
In this language, we define four operator classes (\infer, \proppredict, \propcouple, \emulate, \control) relating state vectors and observations across different systems:
\begin{eqnarray}
\mathcal{I}: \vec{O} = \mathcal{O}[\vec{S}] \rightarrow \hat{\vec{S}}, \quad
\mathcal{P}_{a\to b}: \vec{S}_a \rightarrow \vec{S}_b, \\
\mathcal{E}: (\theta, \vec{S}_0) \rightarrow \vec{S}_{\rm sim}, \quad
\mathcal{C}: (\vec{S},\mathcal{J}) \rightarrow a .
\end{eqnarray}
\end{svgraybox}

The four operator classes cover the full scope of the reviewed work, and are defined as:
\begin{enumerate}
    \item \textbf{Inference} (\infer): Given observations $\vec{O} = \mathcal{O}[\vec{S}]$, recover an estimate of the physical state ($\hat{\vec{S}}$), by approximating the inverse of a known physical forward operator (such as radiative transfer, helioseismic kernels, or instrument point-spread functions). This is the machine-learning analogue to classical physical inversions. It includes point estimates, full posteriors via simulation-based inference, and discrete event labeling.
    \item \textbf{Propagation} ($\mathcal{P}_{a\to b}$): Map one state description to another, either with or without a physical forward observational operator connecting them. This class has two natural sub-cases that dominate the literature: 
    \begin{itemize}
        \item \textbf{Prediction} (\proppredict): Temporal propagation of a given state vector over time, $\vec{S}(t)\to\vec{S}(t+\Delta t)$, mapping directly to standard forecasting.
        \item \textbf{Coupling} (\propcouple): Cross-domain coupling and translation, $\vec{S}_1(t)\to\vec{S}_2(t+\Delta t)$, mapping states across different domains, such as translating EUV coronal images into photospheric magnetograms, or cross-calibrating different telescopes.
    \end{itemize}
    \item \textbf{Emulation} (\emulate): Replace a slow physics-based solver (such as a multi-dimensional MHD or non-LTE radiative transfer code) with a learned surrogate model that takes model parameters or initial conditions as input and returns the simulation output in milliseconds. This includes deterministic surrogates and generative models that sample plausible solar states from a learned distribution $p(\vec{S})$.
    \item \textbf{Control \& Optimization} (\control): Optimal action selection given the current solar state and a scientific objective (e.g., adaptive optics corrections, automated telescope scheduling, or active-learning target selection).
\end{enumerate} 

We note that the key criterion separating \infer\ from \propcouple\ is whether a physical forward operator $\mathcal{O}$ exists such that $\vec{O}=\mathcal{O}(\vec{S})$. If such an operator exists, the ML model approximates $\mathcal{O}^{-1}$ and the task is \infer; otherwise no, the learned mapping encodes inter-subsystem coupling learned from data and the task is \propcouple. Even so, the application under study may consider several of the defined operators at once, and might not present one clear operator in the work. This overlap is not a failure of the taxonomy; rather, it reflects the reality that the Sun is a highly coupled dynamical system.\\
The literature surveyed here clusters into four physical families, each corresponding to a distinct region or problem type: the solar interior (§\ref{sec:interior}); the solar atmospheric state spanning photosphere, chromosphere, transition region, and corona (§\ref{sec:atmosphere}); coherent structures and transient events (§\ref{sec:instability}); observation-space operators covering image restoration, cross-domain translation, representation learning and observatory contro(§\ref{sec:cross_instrument}). Each family is introduced in its physical context in the sections that follow. We finally present a discussion of the applications, and potential path forward in §\ref{sec:conclusion}. 

When discussing any ML application, we ask three questions: 
\begin{enumerate}
    \item What state vector is being inferred, propagated, emulated, or controlled?
    \item What observations constrain the state vector under consideration?
    \item What physical assumptions make the learned operator meaningful? 
\end{enumerate}
This framing keeps the review anchored in solar physics rather than the algorithms themselves. In the solar context, ML approximates operators that are otherwise expensive, ill-posed, or only partially specified. The most mature applications are those where the state vector and operator class are physically explicit. The least secure are those where the output is visually convincing but the implied state vector is poorly defined.
\section{The solar interior}
\label{sec:interior}

The solar interior is inaccessible to direct electromagnetic observation\footnote{We note solar neutrinos as the sole exception, providing a direct, non-electromagnetic probe of the nuclear-burning core. Recent applications of machine learning have successfully lowered detection thresholds in large-scale detectors (e.g., \cite{2025arXiv251219887S}), though a systematic review of these particle physics applications lies beyond the scope of this review.}. All information about its state must be deduced from surface signatures, which probe the convection zone of the Sun. These include acoustic travel times, mode frequencies, and large-scale photospheric magnetic patterns. This is done through forward models whose inversions are computationally expensive and non-unique. Machine learning enters this domain as a tool for constructing tractable approximations to these inverse (\infer) and coupling (\propcouple) operators. The operational fidelity of these inference operators (\infer) remains constrained by the synthetic-to-real domain gap, because synthetic training data rarely captures the full stochastic noise of real telescope observations. It replaces costly pipelines with learned surrogates that can be evaluated rapidly at scale. We organize the interior literature into three interconnected families. The first is deep interior dynamics and physics emulation, targeting the hidden subsurface flow field (\S.~\ref{subsec:interior_deep_emulation}). The second covers latent magnetic states, targeting pre-emergence signatures and geometrically inaccessible far-side magnetic fields (\S.~\ref{subsec:interior_latent_farside}). The third family focuses on global dynamo and solar cycle evolution to capture the long-term magnetic cycle of the internal solar dynamo (\S.~\ref{subsec:interior_global_dynamo}).

\subsection{Deep interior dynamics and physics emulation}
\label{subsec:interior_deep_emulation}

The first family covers the fluid dynamics, kinematic flows, and thermodynamics of the solar interior convection zone. This framework aligns with major collaborative efforts, such as the COFFIES (Consequences of Fields and Flows in the Interior and Exterior of the Sun) framework~\cite{hoeksema2020coffies} and the WHOLESUN project~\cite{wholesun2019synergy}. Both are large-scale, multi-institutional projects aimed at linking subsurface velocity fields with global magnetic structures across the entire Sun. 
\begin{svgraybox}
To model these processes, a representative state vector $\vec{S}_{\rm flow}$ describes the subsurface velocity fields (either in global spherical components $(v_r, v_\theta, v_\phi)$ or local Cartesian components $(v_x, v_y, v_z)$), convective perturbations, and turbulent transport:
\begin{eqnarray}
\vec{S}_{\rm flow} = \{v_r, v_\theta, v_\phi, \delta c, \ldots\},
\end{eqnarray}
where $v_r$, $v_\theta$, and $v_\phi$ represent the radial, meridional, and zonal components of the large-scale velocity fields. Local convective and granular flows are often instead mapped to Cartesian velocities $\vec{v} = (v_x, v_y, v_z)$ and horizontal components $\vec{v}_h = (v_x, v_y)$. The term $\delta c$ denotes perturbations in the local sound speed, acting as a proxy for temperature anomalies in the solar interior. Note that subgrid-scale quantities such as the Reynolds stress tensor $\tau_{ij}$ and acoustic source events $\epsilon$ are not directly retrievable from surface observations; they are simulation-derived quantities used for training labels in the inference-based paradigm.
\end{svgraybox}
The state vector gives rise to different kinds of pressure and gravity waves, which leave periodic signatures in velocities / intensities measured on the solar photosphere. By modeling the photospheric signatures from $\vec{S}_{\rm flow}$, one is able to estimate the inverse mapping ($\mathcal{I}$) to derive the state vector from the observations, where {\ml} has played a significant role.

\subsubsection{Inferring subsurface flows from acoustic diagnostics}
Helioseismology is the primary observational pathway to the solar interior, converting surface Dopplergrams and line-of-sight velocities into wave-travel times and phase shifts, as reviewed in~\cite{2025KosovichevSoPh}. Using these acoustic diagnostics to retrieve subsurface flows is historically a massive computational bottleneck. {\ml} addresses this by learning direct approximations of the inverse operator:
\begin{eqnarray}
\mathcal{I}_{\rm flow}: \vec{D}_{\rm acoustic} = \mathcal{O}_{\rm acoustic}[\vec{S}_{\rm flow}] \rightarrow \hat{\vec{S}}_{\rm flow},
\end{eqnarray}
where $\vec{D}_{\rm acoustic}$ is the vector of observed acoustic diagnostics related to the flow state via the helioseismic observation operator $\mathcal{O}_{\rm acoustic}$.

Two distinct inference problems arise when extracting the solar interior state from these acoustic observations. The first concerns computational scalability. Classical ring-diagram inversions consume over 30 CPU hours per Carrington rotation, creating a bottleneck that prevents near-real-time analysis. Supervised networks trained on ring-diagram mode-fit flow parameters replace this step, yielding surface flow maps in milliseconds. Crucially, they faithfully preserve subtle large-scale flow features---such as equatorial Rossby waves---that aggressive spectral filtering would otherwise suppress~\cite{2019A&A...622A.124A}. 

The second problem concerns acoustic source localization. Weak source signatures are typically overwhelmed by the convective background and resonant p-mode power, making source detection difficult with standard filtering methods. By training a network on radiative-MHD simulations to recognize those patterns, one can recover acoustic emission sites in photospheric Doppler time series and the \textit{network's learned filters} can be reused directly as an interpretable operator~\cite{2021ApJ...915...36B}. A critical extension of this idea is the automated detection of sunquakes, capturing impulsive flare-related seismic signatures that are typically difficult to detect using conventional methods~\cite{2023SoPh..298....4M}. 

\subsubsection{Simulation-backed kinematic reconstruction from intensity}
\begin{svgraybox}
While global models can resolve macro-scale structures, the true multi-scale velocity field within the convective boundary remains physically inaccessible to direct measurement. Here, {\ml} functions strictly as an inference operator ($\mathcal{I}$) to reconstruct these motions not from acoustic waves, but directly from surface intensity patterns:
\begin{eqnarray}
\mathcal{I}_{\rm kine}: \vec{I}_{\rm surf} = \mathcal{O}_{\rm rad}[\vec{S}_{\rm flow}] \rightarrow \hat{\vec{v}},
\end{eqnarray}
where $\vec{I}_{\rm surf}$ represents the emergent surface intensity map related to the flow state through the radiative transfer observation operator $\mathcal{O}_{\rm rad}$, and $\hat{\vec{v}}$ represents either the two-dimensional horizontal velocity field $\vec{v}_h(x, y)$ at the photospheric surface boundary ($\tau_{5000\,\text{\AA}} \approx 1$), or its full three-dimensional depth-dependent extension $\vec{v}(x, y, z)$ into the upper convection zone.
\end{svgraybox}
Surface intensity patterns are a projection of this multi-scale velocity field. Granular and supergranular flows at different depths produce degenerate surface signatures that single-wavelength observations cannot disentangle. This inaccessibility is solved by the \emph{simulation-based inference} paradigm: radiative-MHD simulations provide paired \{surface observable, velocity field\} training examples, making available the hidden-state labels that nature withholds. 

The initial application of this paradigm focused on the surface boundary, where DeepVel~\cite{2018SoPh..293...57T} demonstrated that Convolutional Neural Networks (CNNs)---which extract spatial features by sliding localized filters across an image, assuming that local patterns (like convective granules) look the same regardless of where they appear on the solar disk (translation invariance)---could recover transverse velocity fields $\vec{v}_h$ at granular scales from consecutive surface intensity maps. This successfully surpassed classical Local Correlation Tracking (LCT) and Fourier-based LCT (FLCT) methods by exploiting subtle intensity-morphology correlations that classical optical flow algorithms ignore. Subsequent work progressively extended this boundary mapping into the solar interior. By leveraging temporal and multi-height correlations, operators were trained to recover simultaneously granular and supergranular motions~\cite{2020FrASS...7...25T}, flows within both the quiet Sun and magnetically complex active regions~\cite{2021JSWSC..11....9T, 2025MNRAS.539.2498L}, and height-dependent kinematic signals across multiple photospheric levels~\cite{2026A&A...705A..44K}. The same paradigm extends naturally below the surface: by pairing radiation-MHD simulations with synthetic surface observables, large-scale internal flows at 10~Mm depth can be inferred from just three snapshots separated by one hour, drastically reducing the required input data volume~\cite{2024PASJ...76L..33M}.

Taken together, these results show that the surface and subsurface kinematic state is recoverable from surface observables (up to a certian depth and resolution). The projection of the multi-scale velocity field onto surface observables is inherently degenerate, and this information loss cannot be recovered regardless of simulation quality. Closing the fidelity gap points naturally toward physics-informed approaches that enforce the governing equations directly as constraints, linking this inference framework to the atmospheric reconstruction methods discussed in Section~\ref{sec:atmosphere}.

\subsection{Latent magnetic states: pre-emergence and the far-side}
\label{subsec:interior_latent_farside}

\subsubsection{Subsurface signatures and pre-emergence flux detection}
Flux tubes are plasma structures containing coherent and ordered magnetic field. The rise of magnetic flux tubes through the convection zone alters local thermodynamic properties and generates acoustic travel-time perturbations before the flux emerges at the surface. Detecting these subsurface structures is vital for operational space-weather forecasting, providing early warnings of active region (AR) births. 
\begin{svgraybox}
To formalize this, models target a reduced pre-emergence event state vector,
\begin{eqnarray}
\vec{z}_{\rm em}=\{p_{\rm em}, x_{\rm em}, y_{\rm em}, t_{\rm em}, \Phi_{\rm em}, \ldots\},
\end{eqnarray}
where $p_{\rm em}$ denotes the probability of impending flux emergence, $(x_{\rm em}, y_{\rm em})$ is the estimated emerging spatial coordinate, $t_{\rm em}$ is the lead time until detection, and $\Phi_{\rm em}$ represents the anticipated magnetic flux proxy.
\end{svgraybox}
Rather than executing computationally expensive 3D reconstructions of the magnetic field vector, {\ml} approximates an inference operator. This operator maps photospheric surface observations $\vec{M}_{\rm surf}$ (such as line-of-sight magnetograms or Dopplergrams) directly to estimated event parameters or continuous flux proxies in $\vec{z}_{\rm em}$:
\begin{eqnarray}
\mathcal{I}_{\rm em}: \vec{M}_{\rm surf} = \mathcal{O}_{\rm photo}[\vec{S}_{\rm mag}] \rightarrow \hat{\vec{z}}_{\rm em},
\end{eqnarray}
where $\mathcal{O}_{\rm photo}$ is the photospheric observation operator mapping the subsurface magnetic/dynamic state $\vec{S}_{\rm mag}$ to the surface signatures. Depending on the specific architecture, these models retrieve either discrete components of $\vec{z}_{\rm em}$ (such as the binary probability of emergence $p_{\rm em}$ via patch classification) or spatial projections (such as 2D maps of predicted unsigned magnetic flux $\Phi_{\rm em}(x,y)$).
Pre-emergence detection is fundamentally a weak-signal-in-stochastic-noise problem. A rising magnetic flux tube leaves coherent but faint imprints in surface observables hours before any flux breaches the surface. These subtle signatures are buried beneath the high-amplitude, chaotic convective background of granular and supergranular flows~\cite{2014LRSP...11....3C}.

Instead of classical matched filters that require an analytically specified template, one can derive them directly from the data: by training on labeled pre-emergence and non-emergence line-of-sight magnetograms from SDO/HMI, the machine learning model learns discriminative patterns directly from the data rather than from theory~\cite{2020ApJ...903...27D}. The critical validation is not accuracy on a held-out test set but physical self-consistency. For instance, saliency maps of the trained classifier recover exactly the small-scale dipolar structures predicted by flux-tube rise theory, confirming that the network latches onto the physically causal signature and not onto a proxy correlation. A critical analysis shows that the model incorporates filters sensitive to small-scale, intense magnetic fields and pre-emergence bipolar structures consistent with flux-tube rise theory, confirming that the neural network responds to physically meaningful signatures.

Building on numerical simulations showing that subsurface structures alter acoustic power observed at the photosphere above them~\citep{2011SoPh..268..321H}, modern sequence models—specifically Long Short-Term Memory (LSTM) networks, which are designed to selectively retain or discard information over long time sequences by routing data through gated memory cells—have been designed to predict active region emergence before any visible photospheric manifestation~\citep{2025ApJS..280...64K}. Such models learn the mapping from multi-frequency acoustic power maps to changes in continuum intensity and unsigned magnetic flux. Hence, they are able to successfully forecast the pre-emergence phase-shift signatures and the subsequent localized decrease in continuum intensity that marks active region birth. {\it A pre-emergence model that reports only a binary label makes a weaker scientific contribution than one whose learned features are consistent with physical flux-tube signatures and whose outputs are calibrated probability-of-emergence estimates with uncertainty bounds.}

By extension, the same approach can also probe physical dynamics beyond detection.  Conditioning classifiers on supergranular flow configurations enables quantification of the impact of surface flows on emergence trajectories. These experiments find that quiet-Sun magnetic flux emergence is neither systematically influenced by nor affects the supergranular-scale flows in its vicinity~\cite{2024ApJ...965..186M}. This dual capability—prediction plus causal testing—underscores why model design is critical.

\subsubsection{Far-side active region detection and reconstruction}
Monitoring the Sun's non-visible hemisphere is critical to build synchronic, full-Sun magnetic maps and to anticipate the rotation of active regions onto the Earth-facing disk~\cite{2026SpWea..2404582L}. In this sub-family, the primary state vector is $\vec{S}_{\rm farB}$, representing the magnetic field configuration on the far-side:
\begin{eqnarray}
\vec{S}_{\rm farB} = \{B_r, B_\theta, B_\phi, \Phi_{\rm far}, \ldots\},
\end{eqnarray}
where $B_r$, $B_\theta$, and $B_\phi$ represent the radial, meridional, and zonal components of the magnetic field vector on the far-side, and $\Phi_{\rm far}$ is a scalar activity proxy. In practice, due to the indirect nature of far-side diagnostics, models are often restricted to retrieving simplified projections such as binary active-region detection masks or the radial field component $B_r$.
\begin{svgraybox}
Helioseismic holography can identify large far-side active regions by measuring acoustic phase shifts or travel-time perturbations from multiple-bounce waves ($\delta\vec{\tau}$). These observed seismic phase shifts are related to the far-side magnetic state via a forward helioseismic operator $\mathcal{O}_{\rm seismic}$ such that $\delta\vec{\tau} = \mathcal{O}_{\rm seismic}[\vec{S}_{\rm farB}]$. {\ml} acts as an inference operator ($\mathcal{I}$) to filter this noise and retrieve binary active-region states or magnetic proxies:
\begin{eqnarray}
\mathcal{I}_{\rm farB}: \delta\vec{\tau} = \mathcal{O}_{\rm seismic}[\vec{S}_{\rm farB}] \rightarrow \hat{\vec{S}}_{\rm farB}.
\end{eqnarray}
\end{svgraybox}
Far-side seismic maps suffer from extremely low signal-to-noise ratios because the acoustic waves must travel multiple times across the solar interior, making classical threshold-based detection insensitive to all but the largest active regions. Deep learning models substantially lower this detection floor by functioning as advanced spatial and temporal filters that isolate the weak coherent phase-shift signals from the stochastic acoustic background~\cite{2019A&A...632A..82F}. The natural progression of this inference capability follows an ascending hierarchy of physical richness. First, a binary U-Net establishes the presence or absence of an active region, and is validated against simultaneous STEREO EUV observations~\cite{2021A&A...652A.132B}. Next, attention-enhanced convolutional LSTM (ConvLSTM) networks exploit temporal coherence across successive seismic maps to reduce false alarms~\cite{2022A&A...667A.132B}. Finally, regression networks replace the binary label with continuous synthetic magnetograms generated directly from helioseismic phase shifts~\cite{2024A&A...692A.182B}. This enriches the recovered state from a detection flag to a quantitative estimate of the far-side magnetic flux distribution.

Recent work has begun to close this loop by empirically calibrating the helioseismic phase shift to the unsigned magnetic field using far-side Solar Orbiter/PHI magnetograms~\cite{2025ApJS..278...53H}. Enforcing Hale's polarity law to a pair of Gaussian lobes converts the unsigned far-side signal into a smooth, polarity-resolved magnetogram. Such a scheme avoids unrealistically razor-thin current sheets, approximately conserves flux, and yields far-side magnetograms suitable for assimilation into full-Sun flux-transport and coronal models rather than serving purely as binary activity masks~\cite{2026NatSR..1613110H}.

A complementary observation pathway opens when space missions such as STEREO provide EUV imaging of the non-visible hemisphere. The optically thin coronal emission is controlled by the photospheric magnetic flux in a non-trivial manner. This is exploited to translate EUV imaging observations into quantitative surface magnetograms as a data-driven image-to-image translation problem. Here, deep learning models use a learned surrogate rather than a physical forward operator, mapping the coronal observations $\vec{I}_{\rm EUV}$ to the estimated far-side magnetic state via a cross-domain propagation operator $\mathcal{P}_{{\rm EUV}\to{\rm farB}}$:
\begin{eqnarray}
\mathcal{P}_{{\rm EUV}\to{\rm farB}}: \vec{I}_{\rm EUV} = \mathcal{O}_{\rm EUV}[\vec{S}_{\rm atmos}] \rightarrow \hat{\vec{S}}_{\rm farB},
\end{eqnarray}
where $\mathcal{O}_{\rm EUV}$ is the coronal observation operator mapping the atmospheric state $\vec{S}_{\rm atmos}$ to the observed EUV intensities.
This propagation operator faces a supervision problem: far-side magnetograms are not directly available for supervised training, so the mapping $\mathcal{P}_{{\rm EUV}\to{\rm farB}}$ cannot be trained directly from simultaneous farside observations. The methodological resolution is the \emph{proxy-pair construction} --- training on Earth-facing co-temporal SDO/AIA and SDO/HMI pairs, where the coronal morphology and photospheric field are simultaneously observable, then transferring the learned mapping to the far-side where only coronal morphology exists. The model is validated against withheld Earth-facing ground truth as the active region rotates into view~\cite{2019NatAs...3..397K}. Building on those ideas, new models use EUV 304 and 171\AA\ images together with SFT-predicted magnetic field data from one solar rotation earlier as inputs to reconstruct full-disk farside vector magnetograms. Its main advance is the extension to multispacecraft farside viewing from STEREO-A, STEREO-B, and Solar Orbiter, together with the first comparison of AI-generated farside results against SDO/HMI during near-conjunction intervals and the derivation of active-region magnetic parameters from those reconstructions~\cite{2025ApJS..281...63J}. Together, helioseismic inference and EUV-to-magnetogram propagation establish an operational chain from subsurface acoustic measurements and off-disk coronal images to the photospheric boundary conditions required by global MHD forecasting models. We note, however, that the transition from these models to real-time space-weather forecasting pipelines involves operational challenges (e.g., sensor degradation, telemetry drops, and live data assimilation) that are discussed in detail in the next chapter.

Through the inference ($\mathcal{I}_{\rm farB}$) and cross-domain propagation ($\mathcal{P}_{{\rm EUV}\to{\rm farB}}$) operators, machine learning successfully estimates the far-side magnetic state from noisy or indirect observations. However, because direct measurement is impossible until the active region rotates onto the Earth-facing disk, physical self-consistency becomes the primary metric for algorithmic success. Unconstrained generative models are prone to hallucinating visually convincing features that violate fundamental flux balance. Therefore, the critical next step for these architectures is not merely improving image-to-image accuracy, but embedding strict physical conservation laws, quantifying epistemic uncertainty, and leveraging multi-viewpoint data (e.g., Solar Orbiter) to guarantee the reliability of these synthetic boundary conditions for global forecasting.

\subsection{Global dynamo and solar cycle evolution}
\label{subsec:interior_global_dynamo}

The third physical family scales up from local convective and magnetic features to the macroscopic, long-term behavior of the internal solar dynamo. The solar cycle, which governs the rhythmic 11-year rise and fall of solar activity, is driven by the cyclic regeneration of toroidal and poloidal magnetic fields within the convection zone. Although predictions in this domain rely primarily on surface observations (such as synoptic magnetic maps, sunspot numbers, and $F_{10.7}$ radio flux), they are fundamentally modeling the deep, global magnetohydrodynamic dynamo. 
\begin{svgraybox}
To formalize these macro-scale systems, the internal global dynamo state vector $\vec{S}_{\rm dyn}$ can be defined as:
\begin{eqnarray}
\vec{S}_{\rm dyn} = \{B_{\rm tor}, B_{\rm pol}, u_p, \eta, \ldots\},
\end{eqnarray}
where $B_{\rm tor}$ and $B_{\rm pol}$ represent the internal toroidal and poloidal magnetic field strengths, $u_p$ denotes the meridional circulation velocity, and $\eta$ represents the turbulent magnetic diffusivity.
\end{svgraybox}
Because this deep state is visually inaccessible, machine learning operates in this macroscopic domain in two primary capacities. First, it models the spatial evolution of the photospheric boundary, providing both data-driven forecasting operators ($\mathcal{F}_{\rm SFT}$) for synoptic maps and physics-informed continuous solvers for the governing flux-transport equations. Second, it acts as a temporal propagator ($\mathcal{P}_{\Delta t}$) that forecasts highly compressed historical observations ($\vec{y}_{\rm cycle} = \mathcal{O}_{\rm cycle}[\vec{S}_{\rm surfB}] \in \{\text{SSN}, F_{10.7}\}$) over multi-year baselines. While full inverse data assimilation ($\mathcal{I}$) remains the ultimate theoretical goal for directly inferring the internal $\vec{S}_{\rm dyn}$ state, current architectures are driving advances toward forward-modeling and sequence-prediction pathways.

\subsubsection{Machine learning for surface flux transport and solar cycle modeling}
\label{subsubsec:dynamo_assimilation}

Classical solar cycle predictions rely heavily on Surface Flux Transport (SFT) and kinematic Babcock--Leighton dynamo models to propagate magnetic flux and forecast polar field reversals. Classical SFT describes the advection and diffusion of the radial magnetic field ($B_r$) by differential rotation, meridional circulation, and supergranular dispersal~\cite{yeats_2023_sft}. However, traditional grid-based numerical solvers are computationally expensive and struggle with non-linear stability. {\ml} is increasingly deployed to replace or augment these classical solvers, bifurcating into pure data-driven surrogates and physics-informed continuous solvers.

At the purely data-driven end of the spectrum, image-to-image translation networks based on conditional Generative Adversarial Networks (cGANs) act as direct forecasting operators ($\mathcal{F}_{\rm SFT}$). These models map current synoptic observations directly to future configurations:
\begin{eqnarray}
\mathcal{F}_{\rm SFT}: \vec{B}_{\rm syn}(t) = \mathcal{O}_{\rm syn}[\vec{S}_{\rm surfB}(t)] \rightarrow \hat{\vec{B}}_{\rm syn}(t+\Delta t),
\end{eqnarray}
where $\vec{B}_{\rm syn}(t)$ represents the observed 2D photospheric magnetic flux distribution for a given solar rotation (related to the surface state via the synoptic mapping operator $\mathcal{O}_{\rm syn}$), and $\hat{\vec{B}}_{\rm syn}(t+\Delta t)$ is the predicted global magnetic configuration one full rotation ($\approx 27$ days) later. Rather than explicitly calculating the physical advection-diffusion equations, the cGAN architecture learns the macroscopic transport rules---such as differential rotation and meridional flow---directly from historical data. By utilizing a generative network to synthesize the future map and a discriminator network to enforce pixel-wise structural fidelity, these image-to-image models successfully outperform conventional SFT solvers in global correlation for next-rotation forecasting. By learning the synoptic-map propagation directly from data, they efficiently capture the surface transport of existing magnetic elements~\cite{2025ApJS..278....5J, jeong_2025_radialB_evolution}. However, their predictive horizon---like that of conventional SFT codes---naturally does not encompass the stochastic emergence of new active regions.

To integrate physical fidelity directly into the architectural learning process, Physics-Informed Neural Networks (PINNs) offer a powerful alternative. PINNs embed the governing physical differential equations (such as the advection-diffusion or MHD equations) directly into the network's loss function, penalizing violations of physical laws (e.g., flux conservation) and forcing the model to respect the underlying physics without requiring a conventional grid. Rather than acting as inverse data-assimilation tools, baseline PINN architectures function as highly efficient forward surrogate solvers for the linear SFT equation. By embedding the governing PDE directly into the loss function, the network learns a continuous, mesh-free representation of the flux evolution without requiring explicit spatial discretization or numerical time-stepping, demonstrating better flux conservation and mesh independence compared to classical explicit schemes~\cite{jithu_2024_sft}.

The deeper advantage of this differentiable framework emerges when scaled to a non-linear Babcock--Leighton dynamo regime. Global cycle models depend heavily on the non-linear feedback mechanisms that govern the conversion of toroidal flux into poloidal flux. PINN-based forward solvers robustly capture these complex non-linearities achieving superior numerical stability and flux conservation compared to traditional upwind schemes~\cite{2026ApJ..1000...79A}. While current implementations operate with fixed transport parameters, establishing these continuous differentiable solvers lays the critical groundwork for future multiparameter data assimilation and the simultaneous inversion of the internal dynamo state.

\subsubsection{Long-term time-series and solar cycle forecasting}
Because the internal dynamo state remains inaccessible to direct observation, most machine-learning approaches forecast low-dimensional observables that encode information about the global magnetic cycle, such as sunspot number, $F_{10.7}$ flux, polar field strength, or synoptic magnetic harmonics. 
\begin{svgraybox}
The most widespread application in this family is the direct forecasting of cycle-related observables using historical activity indices (including retrospective predictions for Cycle 24 and forward forecasts for Cycles 25 and 26). Here, machine learning acts as a temporal propagation operator, mapping a history of past observations to a future forecast:
\begin{eqnarray}
\mathcal{P}_{\Delta t}: \vec{y}_{\rm cycle}(t) = \mathcal{O}_{\rm cycle}[\vec{S}_{\rm surfB}(t)] \rightarrow \hat{\vec{y}}_{\rm cycle}(t + \Delta t),
\end{eqnarray}
where $\vec{y}_{\rm cycle}$ represents the vector of observed cycle activity indices (e.g., SSN, $F_{10.7}$) through the observation operator $\mathcal{O}_{\rm cycle}$, and $\Delta t$ ranges from several months to a full solar cycle (\text{approx.}~11~yr). These models primarily predict cycle amplitude, peak timing, and overall temporal evolution.
\end{svgraybox}
A physical motivation for sequence-based architectures is that solar activity exhibits memory on multiple timescales: the Hale magnetic cycle ($\sim$22~yr) modulates the 11-year sunspot envelope, meridional circulation imposes a multi-year transport delay on polar field buildup, and stochastic active region emergence introduces cycle-to-cycle scatter on shorter timescales~\cite{2014LRSP...11....3C}. From a dynamo perspective, forecasting the sunspot number (SSN) time-series can be viewed as a highly compressed, scalar reduction of the global photospheric boundary state ($\vec{S}_{\rm surfB} \to \text{SSN}$), serving as an accessible proxy to constrain and predict the unobservable internal dynamo state. Early baseline attempts using simple, static feed-forward networks~\cite{1995ApJ...444..916C, 1996AnGeo..14...20F, 2009JASTP..71..569U} demonstrated proof-of-concept forecasting, although their feed-forward architectures lacked explicit mechanisms for retaining information across long temporal sequences. In contrast, Long Short-Term Memory (LSTM) networks address this directly by selectively retaining or discarding information across sequences of arbitrary length. LSTM architectures trained on historical sunspot indices typically outperform classical regression, with benchmarks across non-deep, deep, and ensemble methods establishing the relative skill of each class~\cite{2021RAA....21...12W,2021RAA....21..184L,2022AdSpR..69..798P,2023SoPh..298...50P,2022arXiv220305757D}.

The prediction target extends beyond scalar sunspot number. Machine-learning methods have been applied to daily $F_{10.7}$ radio-flux indices~\cite{2009RAA.....9..694H}, to the planetary geomagnetic activity index $Ap$~\cite{2018SpWea..16.1424O}, to the sunspot butterfly diagram itself~\cite{2019SoPh..294...24C}, and to spherical-harmonic representations of synoptic magnetic maps~\cite{2024ScChD..67.2460Z}, extending the forecasting problem from scalar activity proxies to the evolution of the large-scale magnetic field itself. In the same vein, machine-learning analyses of extended historical records have identified 120-year oscillatory variations and forecast future extended solar minima~\cite{2021AdSpR..68.1485V,2024AdSpR..74.4173R}.

The methodological progression in this family reflects three distinct forecasting objectives. First, capturing the multi-scale temporal structure requires models capable of resolving both short- and long-term dependencies simultaneously~\cite{benson_2020_solarcycle,su2024_solarcycle}. Second, resolving long-range dependencies to capture sub-cycle features over the full historical sequence requires transformer-based architectures or multivariate models~\cite{2024ApJ...969..120C,2024SoPh..299..117R}. Third, moving beyond point forecasts toward probabilistic forecasting provides more actionable output; Bayesian forecasting architectures that constrain uncertainty bounds on cycle amplitude and timing~\cite{2022MNRAS.515.5062B} offer a more operationally relevant product than deterministic forecasts, because forecast uncertainty can be explicitly incorporated into planning and risk assessment. 

Machine learning has established that the solar cycle amplitude, peak timing, and sub-cycle fine structure are all partially predictable from historical activity indices. However, true operational forecasting remains challenging; as noted in recent comprehensive reviews~\cite{2026arXiv260416183K}, many pure ML models struggled to accurately predict Solar Cycle 25's strength ahead of time. A fundamental limitation of purely data-driven forecasting is the intrinsically small sample size available for supervised learning. Even with historical sunspot records spanning centuries, the availability of only approx. two dozen solar cycles severely limits statistical power and increases overfitting risk. Furthermore, current machine-learning models primarily propagate observable surface proxies forward in time, rather than addressing the central physical challenge: inferring the latent dynamo state $\vec{S}_{\rm dyn}$ from incomplete observations. Therefore, the path toward operationally useful forecasting ultimately lies in the hybrid data-assimilation frameworks described in Section~\ref{subsubsec:dynamo_assimilation}, combining the temporal sequence modeling of deep learning with the physical constraints of continuous differential solvers.
\section{The solar atmosphere: from surface to corona}
\label{sec:atmosphere}

The solar atmosphere is a highly tenuous plasma where magnetic field dynamics increasingly dominate the thermodynamics with height. From the photosphere onwards, the atmosphere experiences a near-exponential drop in density across $\approx10$ orders of magnitude. Alongside this, we observe a dramatic rise in temperature from $\approx5500$ K to $\geq10^6$ K. The plasma also undergoes varying stages of ionization from the photosphere to corona, where it is fully ionized. Furthermore, the magnetic field starts playing an increasingly larger role in the solar atmosphere, adding to the complexity of processes in the system.

Structurally, the solar atmosphere is divided into four main layers: the photosphere, the chromosphere, the transition region, and the corona. Just above the photosphere lies the chromosphere, which typically consists of partially ionized plasma with temperatures rising from a minimum of $\approx 4.4 \times 10^3$ K up to $\approx 2 \times 10^4$ K. Above this, the temperature rapidly climbs up in a geometrically very thin layer to $\approx10^6$ K, which we call the transition region. Owing to the sharp reduction in gas pressure, the magnetic field starts to dominate the dynamics of this region. Above the transition region lies the solar corona, where temperature can go higher than $10^6$ K. Typically, the dynamics in this part of the atmosphere is dictated by the magnetic field, including the morphology, strength, etc. In general, the entire atmosphere is quite tightly coupled, with mass and energy transfer from bottom to the top and vice-versa.

The predominant mode of observations in this region is through spectral lines, which form at different characteristic temperatures and pressures~\footnote{We also note that continuum is also studied.}. The observation mode is either through spectral band-passes covering several spectral lines, purely spectral, or spectropolarimetric. The observations can be either taken on ground or in space, with each having its own set of complications and issues. Since a wealth of spectral lines form across various conditions in the solar atmosphere, we are able to generate a sense of the typical plasma state in the atmosphere. In particular, this wealth of data means a strong contribution of {\ml} techniques in understanding the plasma state in the atmosphere at a wide range of spatial, temporal, and energy scales.  \\
The state vector is defined as:
\begin{eqnarray}
\vec{S}_{\rm atmos}=\{T,\chi_i,n_e,\vec{v},\vec{B}, \xi, \ldots\}(x,y,z,t),
\label{eqn:statevec_atmos}
\end{eqnarray}
where $T$ represents temperature, $\chi_i$ the composition, $n_e$ the electron number density, $\vec{v}$ the velocity vector, $\vec{B}$ the magnetic field vector, and $\xi$ the microturbulence, all as a function of spatial coordinates and time, respectively. Sometimes it is non-trivial to work in geometric height, and instead prefer to work in terms of \emph{optical depth} ($\tau$), which is defined as $$d\tau = \alpha_\nu(s)ds.$$ Here, $\alpha_\nu$ would be the absorption coefficient of the plasma, which depends on the radiation properties, composition, density, etc. and $s$ is the line of sight length, which would be some combination of path length from $x,y,z$. The main observables from this state vector would be spectral \& spectro polarimetric observations, narrow band filtergrams across different wavelengths, and radio signatures.

\subsection{Inference: thermodynamic and plasma state of the atmosphere}
\label{subsec:atm_thermo}
\begin{figure}
    \centering
    \includegraphics[width=\linewidth]{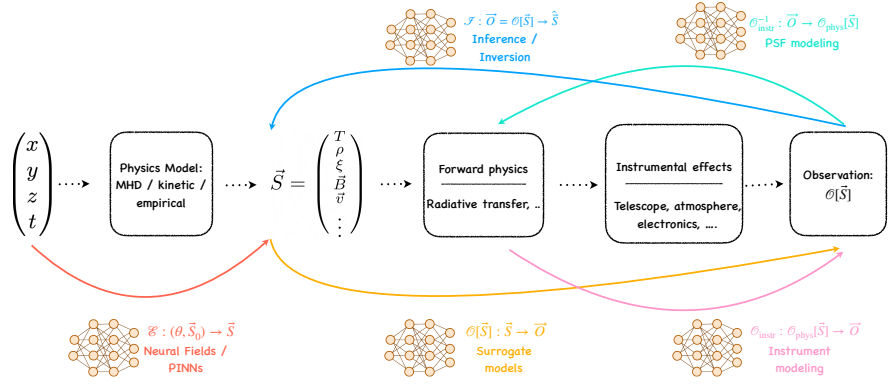}
    \caption{Several ways of using {\ml} to perform inference of the solar state vector given observations. These include performing a direct inversion given the observations, or model the forward process in a data-driven manner.}
    \label{fig:forward_op}
\end{figure}

The observations in this section are either spectral, spectropolarimetric or imaging observations of the atmosphere. Given the observations ($ \mathcal{O}[\vec{S}_{\rm atmos}]$), we seek to obtain $\vec{S}_{\rm atmos}$. This is depicted pictorially in Fig.~\ref{fig:forward_op}. We note that such an inference can be accomplished in various ways: through development of a direct inference (\infer) method, or through the development of surrogate models ($\mathcal{O}$), or through the development of emulators (\emulate) coupled with existing inversion methods. 

In the lower solar atmosphere, the path length for photons is short when compared to the line of sight length, so effectively the light seen is formed in optically thick conditions. Effective inference (\infer) in optically thick regimes requires full radiative-plasma modeling, as simpler approximations often fail to capture vertical stratifications. However, higher up in the atmosphere, the path length for photons is long, rendering the radiation optically thin. In such conditions, $\vec{S}_{\rm atmos}$ can be effectively reduced to a few free parameters, since different parameters get decoupled from radiation.
\subsubsection{Optically thick plasmas}
\label{subsubsec:optthickplasma}
A Stokes measurement encodes the full polarimetric state of light into four quantities: total intensity ($I$), linear polarization along two axes ($Q$, $U$), and circular polarization ($V$) --- making it the primary observational diagnostic of magnetic field strength and orientation in the solar atmosphere.

In the photosphere and chromosphere, much of the information about the atmosphere is encoded in spectral-line profiles and the Stokes vectors. Recovering these physical parameters from spectra constitutes a fundamental inversion task (\infer). A key observational challenge here is that polarimetric signals are extremely weak. They are easily corrupted by instrumental noise, polarization crosstalk, and atmospheric turbulence (seeing). Consequently, advanced spatial and spectral denoising is required before physical parameters can be reliably recovered. These spectral lines form in optically thick conditions, making the line formation highly complex. A compact lower-atmosphere state vector may be written as:
\begin{eqnarray}
\vec{S}_{\rm atmos, thick}=\{T,\rho,n_e,\vec{v},\vec{B},\chi_i,\ldots\}(x,y,z,t),
\end{eqnarray}
where $T$ represents temperature, $\rho$ the mass density, $n_e$ the electron density, $\vec{v}$ the velocity field, $\vec{B}$ the magnetic field vector, and $\chi_i$ the chemical composition. The most straightforward estimation of $\vec{S}_{\rm atmos, thick}$ is to estimate the inversion operator such that:
\begin{equation}
    \mathcal{I}: \mathcal{O}[\vec{S}_{\rm atmos, thick}]\rightarrow \vec{S}_{\rm atmos, thick}.
\end{equation}

Early neural-network inversions established the foundation for modern techniques by demonstrating learned mappings from Stokes profiles to thermodynamic parameters and depth-dependent stratifications~\cite{carroll2001_ann,carroll2008_strat}. Systematic tests also showed improved stability and noise tolerance relative to classical iterative schemes~\cite{2005ApJ...621..545S}.

Because inversions rely heavily on repeated forward modeling, a critical application of {\ml} has been accelerating the forward radiative-transfer calculations themselves. Learned radiative-transfer models act as fast emulation operators (\emulate) to generate synthetic Stokes profiles directly from 1D atmospheric states, bypassing traditional synthesis codes~\cite{2024RMxAA..60..343R_stokessurrogate,2025A&A...703A..55A_stokes_andres}. Beyond end-to-end profile synthesis, a highly practical strategy is learning surrogates for expensive intermediate quantities, such as non-LTE departure coefficients. Utilizing 3D convolutional neural networks (CNNs) and Graph Neural Networks (GNNs)---which represent physical quantities as nodes on a graph and map their couplings along connections (edges) to model non-local interactions like multi-line radiative transport---these surrogates bypass the traditional statistical equilibrium solver, achieving order-$10^3$ acceleration while preserving chromospheric structure~\cite{2022A&A...658A.182C_radtransfer,2022ApJ...928..101V_gnn_hazel,2026arXiv260509543V}. Alternatively, the statistical equilibrium solver can be reframed as an active control task (\control). Policy-Based Radiative Transfer utilizes model-free reinforcement learning policies to discover the depth-dependent source function dynamically, eliminating the need for approximate lambda operators~\cite{2025arXiv250415679P}.

Returning to the inverse problem (\infer), some retrieval systems leverage these speedups through hybrid architectures by providing high-quality initializations for iterative inversions, accelerating convergence while retaining physics-based refinement~\cite{2021A&A...651A..31G}. When fully supervised mapping is employed, training sets are typically constructed in two ways depending on the specific task and the available realism: using synthetic data from numerical simulations degraded by instrumental effects, or using real observations paired with classical inversion results~\cite{2019A&A...626A.102A_stokesInv, 2020A&A...644A.129M, 2020ApJ...894...70L_simbasedinf, yang_2024_specpol1, yang_2025_specpol2, 2019ApJ...875L..18S_iris2_simbasedinf, 2026ApJ...997..229S_iris2p_simbasedinf}. Feed-forward architectures like CNNs and Transformers~\cite{campbell_2026_transformers_specPol}---which use self-attention mechanisms to weigh the relationships between different parts of an input (such as different spectral channels or pixels) regardless of their physical distance---excel at these fast mappings. By drastically reducing inference time, these models enable continuous, large field-of-view studies that turn single-pixel physical conclusions into population statistics. Examples include recovering chromospheric properties of Ca~II~K fibrils across an entire field of view~\cite{2020A&A...637A...1K_caII_fibrils}, or mapping the spatio-temporal distribution of chromospheric heating across a plage region at high cadence~\cite{2022A&A...664A...8M_plage_heating}. Such tasks were previously computationally intractable with classical pixel-by-pixel inversions. Beyond scaling up single-instrument inversions, this distillation paradigm can even bridge disparate observational domains: Visual Transformers have been deployed to translate broadband imaging and magnetograms (SDO/AIA and HMI) directly into chromospheric thermodynamic states, using IRIS$^2$ spectral inversions as the ground-truth target~\cite{2026arXiv260421184S}. A fundamental caveat of these supervised approaches is that their physical fidelity is strictly bounded by the realism of the training set. Synthetic models often fail to capture the full complexity of real observations, such as stochastic noise and instrumental artifacts. Furthermore, a new model must typically be trained from scratch for each new instrument, which remains computationally expensive and time-consuming.

To circumvent the biases of simulation-based training sets, a separate class of architectures shifts away from feed-forward prediction toward per-observation optimization using PINNs and Neural Fields. In these representations, the state vector is defined such as: $$\vec{S}_{\rm atmos, thick} = \mathbf{NN} (x,y,z,t),$$ where $\mathbf{NN}$ is a neural network representation of the state vector. The physics forward model is assumed to be a smooth, differentiable framework, and the problem is defined as:
\begin{equation}
\mathbf{NN}: \{x,y,z,t\} \rightarrow \mathcal{O}[\mathbf{NN}(x,y,z,t)].
\end{equation}
These methods embed the radiative-transfer equations directly into the optimization loss function~\cite{2025ApJ...985L...7J_stokes_me,2025arXiv250709430L_stokes_milne}. This idea of using a neural network to represent the atmosphere as a continuous field can be very powerful to enforce spatio-temporal consistency across the atmosphere providing results that are not only visually convincing but also physically consistent and robust against noise and ambiguities~\cite{2025A&A...693A.170D_neuralfields_stokes_wfa}. Unlike fast surrogates, these methods require optimization at inference time, but they guarantee that the retrieved parameters strictly enforce physical and spatial constraints.

Finally, because spectropolarimetric inversion is mathematically non-identifiable under limited observables, deterministic point estimates are physically incomplete and the uncertainty of the inferred state vector becomes vital for scientific interpretation~\cite{2023SoPh..298...98M_stokes}. Probabilistic models explicitly target this ill-posedness. In highly dynamic flare conditions, where standard assumptions fail, invertible neural networks (INNs) provide a probabilistic route by jointly learning forward and inverse mappings, successfully recovering thermodynamic structure from complex RADYN-simulated profiles like H$\alpha$ and Ca II 8542~\cite{2019ApJ...873..128O_simbasedinf}. Approximate Bayesian inference can also be framed as a Doppler imaging problem, where neural networks learn to map stellar line profiles to surface abundance maps while capturing posterior uncertainty~\cite{2022A&A...658A..16A_bayesdoppler}. To obtain the full posterior distribution of $\vec{S}_{\rm atmos, thick}$, Normalizing Flows ---which learn complex probability distributions by transforming a simple distribution (like a Gaussian) through a sequence of invertible, differentiable steps—offer amortized posterior estimation, shifting the computational cost entirely to training while enabling rapid, mathematically calibrated Bayesian inference at test time~\cite{2022A&A...659A.165D_normflows_stokes} for individual pixels or even performing Bayesian inference across an entire 2D field-of-view using flow-matching techniques~\citep{2026arXiv260613004A_stokesflow}.
\subsubsection{Optically thin plasmas}
\label{subsubsec:optthinplasma}

With increasing height and reducing density, radiation forms in optically thin conditions. In the transition region and the corona, the excitation of the upper atomic levels can be approximated as collision-dominated, while de-excitation is predominantly due to spontaneous emission~\cite{2004psci.book.....A}. This simplification decouples the radiation field from the local plasma state, allowing us to express the emission purely in terms of temperature, electron density, and ion properties. Consequently, our state vector reduces to $\vec{S}_{\rm atmos, thin}=\{T,\chi_i,n_e,\vec{v}\}(x,y,z,t)$, where $T$ represents temperature, $\chi_i$ the composition or ionization fraction, $n_e$ the electron density, and $\vec{v}$ the velocity field. 

For spectral line emission, this discrete state vector must be integrated along the line of sight. In optically thin conditions, the observed intensity $I$ is classically written as:
\begin{eqnarray}
    I \propto \int n_e^2 G(T) ds,
    \label{eqn:goft}
\end{eqnarray}
where $G(T)$ is the contribution function of the spectral line of interest, encapsulating the relevant atomic physics (elemental abundance, ionization fraction, and excitation rates). In a more general setting that accounts for macroscopic plasma motions (Doppler shifts), this can be rewritten as~\cite{2019ApJ...882...13C_vdem}:
\begin{eqnarray}
    I \propto \int_{v}\int_{T} \mathcal{D}(T, v) \mathcal{G}(T, v) dT dv,
    \label{eqn:dem}
\end{eqnarray}
where $\mathcal{D}(T, v)$ is the \emph{Differential Emission Measure} (DEM), a macroscopic measure of the plasma density distribution across different temperatures and velocities along the line of sight. If the observations permit, these integrals can be further expanded to disambiguate specific chemical compositions. Because $\mathcal{D}(T,v)$ is a property of the plasma rather than of any particular instrument passband, it provides a portable, instrument-agnostic description of the thermal state --- making it the natural quantity for cross-mission comparison and multi-wavelength synthesis.

Because the line-of-sight integration destroys 3D spatial information, multiple different thermal distributions can produce the identical observed intensity. Classically, this is solved using linear solvers stabilized by regularization~\cite{2012A&A...539A.146H_dem,2015ApJ...807..143C_dem,2019ApJ...882...13C_vdem,2020ApJ...905...17P_dem,2025A&A...695A.125Y_dem}. Machine learning transforms this by deploying learned inference operators to instantly map multi-wavelength intensities to DEM distributions. However, because of the inherent mathematical degeneracy of the integral, deterministic point-estimate networks are often physically insufficient. The most robust formulations explicitly solve for the full posterior distribution~\cite{1998ApJ...503..450K_dem}, quantifying the range of thermal structures consistent with the observed intensities.

Beyond standard DEM inversions, machine learning operators can disentangle overlapping physical regimes that violate the pure optically thin assumption. Very frequently, cool chromospheric plasma (e.g., filaments) is suspended in the corona, absorbing underlying EUV radiation. Deep learning techniques act as inference operators ($\mathcal{I}$) to mathematically separate the cool, optically thick absorbing component from the hotter, optically thin background plasma~\cite{2026A&A...707A.237M_hot_cool_disamb}. 

Finally, observing the off-limb solar corona (such as in white-light coronagraphy) presents a distinct geometrical inverse problem. Here, photons undergo Thomson scattering off free electrons, making the direct target the three-dimensional electron density. This 3D tomographic reconstruction can be achieved by using neural field representations and solving the forward scattering problem as a differentiable operator~\citep{2023SoPh..298..135A} or as an image translation problem from synoptic disk observations to electron density maps~\cite{2021ApJ...920L..30J_density_coronograph}. 

\subsection{Three-dimensional magnetic-field extrapolation}
\label{subsec:atm_mag_inference}

A very specific subsection of the atmospheric problem is dedicated to estimating the three-dimensional magnetic field across the upper solar atmosphere. Significant dynamics in the solar atmosphere are controlled by the magnetic field, which couples mass and energy flows between the lower and upper atmosphere. Because direct 3D measurement of the coronal field remains observationally challenging, these volume fields must be derived from boundary estimates. We define this lower boundary state as:
\begin{eqnarray}
\vec{S}_{\rm surfB}=\{\vec{B}_{\rm phot}\},
\label{eqn:b_state}
\end{eqnarray}
where $\vec{B}_{\rm phot}$ represents the vector magnetic field at the photosphere. The extraction and refinement of this state-of-the-art boundary condition from raw observations---including cross-instrument harmonization (e.g., SDO/HMI to Hinode/SOT), spatial super-resolution, and far-side reconstruction---are structured as observation-space operators ($\mathcal{P}_{1 \to 2}$) and global temporal propagators, which are discussed extensively in Sections \ref{sec:cross_instrument} and \ref{sec:interior}, respectively.

Assuming a reliable surface boundary state $\vec{S}_{\rm surfB}$ has been retrieved, one seeks the 3D coronal field:
\begin{eqnarray}
\vec{B}_{\rm cor}=\mathcal{P}_{{\rm z =0}\rightarrow{\rm z}}\left[\vec{S}_{\rm surfB}\right].
\end{eqnarray}
This must satisfy physical constraints such as $\nabla\cdot\vec{B}=0$ and, in force-free models, $\nabla\times\vec{B}=\alpha\vec{B}$. Hence, this step is not an inference problem (\infer) from local emission, but a spatial propagation problem (\propcouple). Potential-field models assume zero currents ($\alpha=0$); linear force-free models allow a spatially uniform current helicity ($\alpha = \mathrm{const}$); and nonlinear force-free (NLFFF) models permit spatially varying $\alpha$, making them the most physically realistic but also the most computationally demanding.

{\ml} addresses this extrapolation problem in two distinct ways. The first is solution fitting: a continuous neural representation is optimized for a particular active region or boundary condition while penalizing violations of the governing Maxwell equations. PINNs provide this grid-free formulation, successfully achieving force-free coronal-field extrapolations~\cite{jarolim_2023_pinnextrap,zhang_2024_pinnforcefree}. Because the photospheric field is not itself fully force-free, applying multi-height constraints (incorporating chromospheric fields) significantly improves the physical validity of these PINN extrapolations~\cite{jarolim_2024_pinnchrom}. Additionally, generative approaches provide a probabilistic route to mapping boundary information into a coronal volume; for instance, denoising diffusion models operating on spherical harmonics can synthesize physically realistic 3D coronal magnetic fields across the solar cycle~\cite{2025arXiv251001441D}.

The second route is operator learning. Instead of optimizing a new neural field from scratch for each active region, one learns the mapping from a boundary magnetogram directly to the 3D coronal field across a massive training set. We write this as an emulation problem:  
\begin{eqnarray}
\vec{B}_{\rm cor}=\mathcal{E}_{{\rm z =0}}\left[\vec{S}_{\rm surfB}\right].
\end{eqnarray}
Using architectures such as Fourier Neural Operators (FNOs) and Physics-Informed Neural Operators (PINOs)~\cite{li_2020_FNO,jeon_2025_realtime3dnlff,yang_2026_neuralop}, the network provides a direct, single-pass estimation of the full 3D field. FNOs learn mappings between continuous infinite-dimensional function spaces (such as 2D boundary fields to 3D volume fields) directly in the frequency domain, allowing them to remain independent of the grid resolution. While this functions as a highly accelerated emulator ($\mathcal{E}$) of classical extrapolation codes at inference time, it comes with a fundamental limitation: the learned operator strictly inherits the assumptions, errors, and systematic biases of the specific numerical NLFFF solvers used to generate its training targets.

Finally, global coronal-field modeling introduces an additional layer of complexity because the spatial propagation ($\mathcal{P}_{{\rm z =0}\rightarrow{\rm z}}$) requires a synchronic boundary condition that includes the unobserved hemisphere. As discussed in Section~\ref{sec:interior}, AI-generated far-side magnetograms~\cite{jeong_2020_extrap_farside} close this loop. This represents the ultimate coupled inference chain: an inference operator ($\mathcal{I}$) reconstructs the hidden far-side boundary, a spatial propagation operator ($\mathcal{P}$) extrapolates the global boundary into the 3D corona, and the resulting coronal volume serves as the physical input for downstream space-weather forecasting models.

\subsection{Evolution of the atmospheric magnetic state}
\label{subsec:atm_evolution}

The atmosphere is not a sequence of independent images. At active-region scales, its state evolves dynamically under flux emergence, cancellation, shearing, rotation, and reconnection. {\ml} models therefore act as temporal propagation operators ($\mathcal{P}$), bridging the gap between static extrapolations and fully dynamic MHD:
\begin{eqnarray}
\vec{S}(t+\Delta t) = \mathcal{P}_{t\rightarrow t+\Delta t}[\vec{S}(t)].
\end{eqnarray}
At these localized scales, learned operators predict the short-term evolution of the photospheric magnetic field from sequences of earlier magnetograms~\cite{bai_2021_predphotosphericfield}. Rather than simulating the slow advection of the global solar cycle (which is governed by the interior dynamo; see Section~\ref{sec:interior}), these atmospheric propagation operators focus on the rapid topological changes that precede energetic events.

To support global atmospheric evolution, handling missing observational data is critical. Earth-view magnetograms do not observe the full Sun simultaneously, yet global coronal and heliospheric models require a synchronic boundary condition. One approach is to infer far-side magnetic information from multi-spacecraft EUV observations~\cite{2025ApJS..281...63J}. Another is to generate magnetogram sequences directly from EUV image sequences, treating the dynamic EUV corona as an indirect tracer of the evolving magnetic lower boundary~\cite{2022ApJS..262...45S_aia2magentogram}. These methods couple atmospheric layers in a practical sense: coronal emission is used to continually update a hidden photospheric magnetic state that drives subsequent coronal or heliospheric modeling.
\section{Coherent magnetic structures and transient events}
\label{sec:instability}
Solar observations reveal ordered magnetic and thermodynamic structures spanning multiple spatiotemporal scales, though their apparent morphology is heavily biased by instrumental sensitivity. These structures may be in an \emph{unstable equilibrium}, and may become unstable to release mass and energy into the atmosphere. 
\begin{figure}
    \centering
    \includegraphics[width=\linewidth]{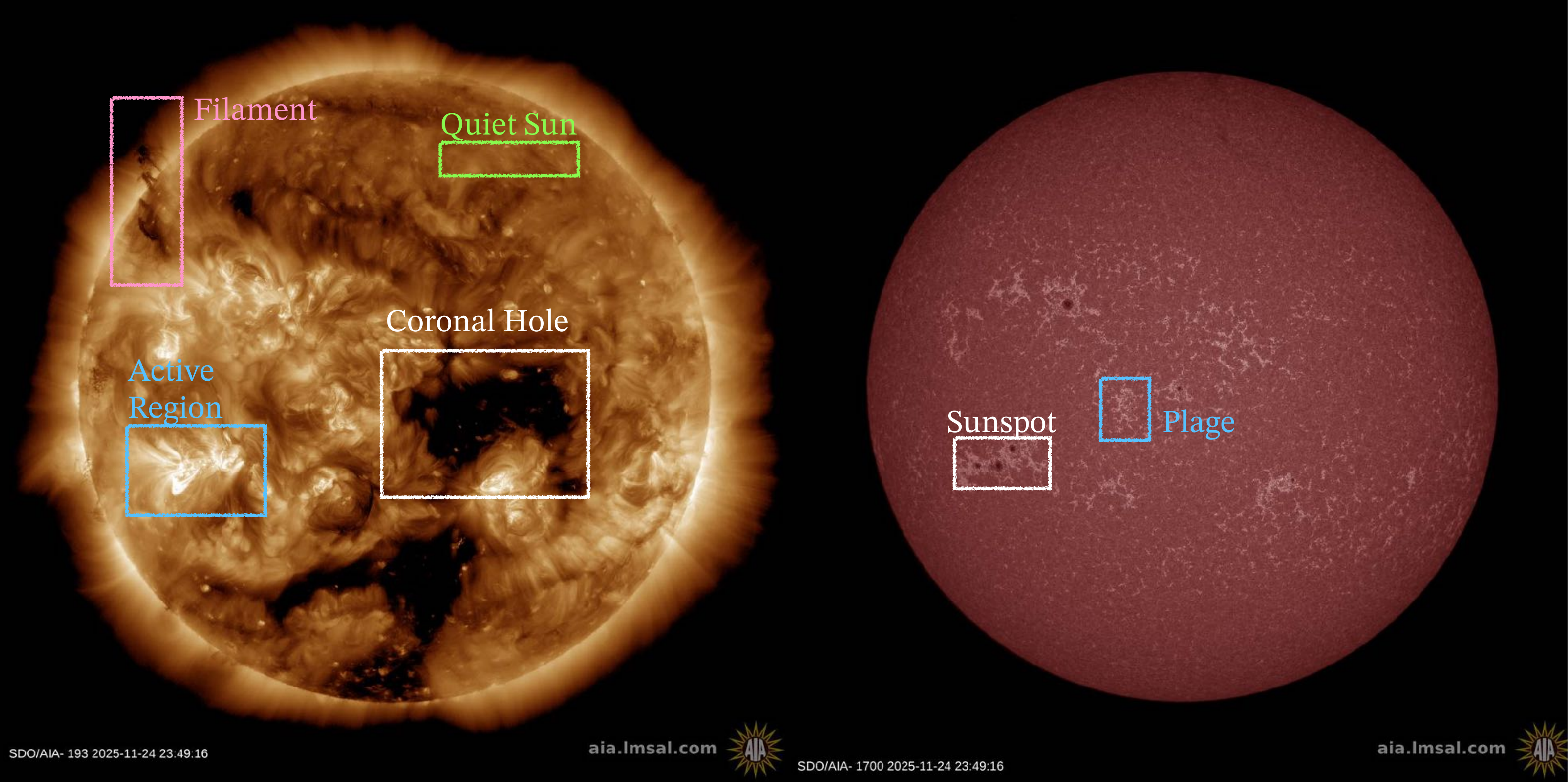}
    \caption{Solar observations in AIA at 193~{\AA} in the left panel, and AIA 1700~{\AA} in the right panel. Several features in the images are marked in both the images. Credit: SDO/AIA team from LMSAL.}
    \label{fig:aia_features}
\end{figure}
As illustrated in Fig.~\ref{fig:aia_features}, distinct atmospheric layers exhibit different morphological features (e.g., coronal holes, filaments) driven by variations in their emission measure profiles. In the AIA 193~{\AA} image, we have marked several regions: coronal holes, active regions, filaments, and quiet Sun. The differences in the intensities may be understood through the formalism of DEM ($\mathcal{D}(T, v)$; from Eq.~\ref{eqn:dem}). We have seen in Eq.~\ref{eqn:dem} that for optically thin plasmas, the intensity is a function of both $\mathcal{D}(T, v)$ and $\mathcal{G}(T, v)$. Coronal holes represent regions of open magnetic flux characterized by both lower plasma density and reduced temperatures compared to the quiet Sun. This dramatic drop in emission measure ($n_e^2$) across temperatures sampled by the AIA 193~{\AA} passband is why they appear as dark regions in the image~\cite{aiaresponse}. On the other hand, active regions contain enough plasma in the AIA 193~{\AA} temperature sensitivity, and hence appear bright. These structures are visible above a rather persistent background, which we call the quiet Sun. The filaments we see in this image present an interesting physical case --- visually, one would think they have properties similar to coronal holes. However, multi-wavelength observations tell us that filaments are cold plasma suspended above hotter plasma below. They appear dark because background EUV emission is absorbed by dense, cool plasma via photoionization continua, not because there is a lack of emitters. In the right panel of the same figure, we display an image of AIA 1700~{\AA}, where we have marked two important regions --- sunspots and plage. In this image, the DEM formalism fails since the observation is in the optically thick regime. However, through the aid of modeling and multiwavelength observations, we can say that plage corresponds to concentrated magnetic field which gives rise to enhanced emission. In the same image, we see the darker sunspots. One might think the magnetic field here is weaker, but the exact opposite is true. The concentrated magnetic field in a sunspot suppresses local convection. This blockage prevents energy transport from the interior, lowering the surface temperature and the observed intensity.

Apart from the features marked in this image, features and activity on the Sun occur at a wide range of scales, each of which couple across different parts of the atmosphere. Furthermore, some of these structures couple not just across the layers of the atmosphere, but have an impact through the whole heliosphere. In this section, we shall consider literature that work with morphological structures and features across different length, time and energy scales. 

\subsection{Structure identification}
\label{subsec:trans_segmentation}
We first focus on algorithms developed for feature identification. The state vector for \emph{any particular feature} is the same as those of the solar atmosphere, as enumerated in Eq.~\ref{eqn:statevec_atmos}. For the state vector $\vec{S}_{\rm feat}$ of feature, the feature itself is defined as: 
\begin{equation}
    \mathcal{F}_{\rm feat} = \{(x,y,z,t) \in \mathbb{R}^4 \mid  \vec{S}_{\rm atmos} (x,y,z,t) = \vec{S}_{\rm feat}\},
    \label{eqn:features_segment}
\end{equation}
where we define a feature ($\mathcal{F}_{\rm feat}$) as a set of points in space and time where the state vector is consistent with the state vector defining the feature. We operate in the observation space, so feature identification translates to an inference operator (\infer) such that:
\begin{eqnarray}
\mathcal{I}_{\rm feat} : \vec{O} = \mathcal{O}[\vec{S}_{\rm atmos}, \vec{S}_{\rm feat}] \rightarrow \mathcal{F}_{\rm feat}.
\label{eqn:infer_feat}
\end{eqnarray}

\subsubsection{Spatial Boundary Extraction}
The oldest observations of the Sun correspond to white-light images of the photosphere, where sunspots are seen very clearly. Hence, it is one of the oldest data-driven problems in solar physics, dating back centuries to recorded sunspot diagrams~\cite{2007AdSpR..40..929V}. The foundational challenge is that the boundary between umbra, penumbra, and quiet Sun is not a sharp physical interface but a continuous gradient in $\vec{S}_{\rm atmos}$. Early work focused on automated sunspot group classification based on feedforward neural networks~\cite{2008SoPh..248..277C}. Better representations of these segmentation masks were developed by coupling morphological operations to generate a more complete $\mathcal{F}_{\rm sunspot}$ ~\cite{2015JSWSC...5A..15A}.

In all these tasks the ground truth, on which the model is trained, is very critical. Hence, it is imperative to have schemes for a well-posed definition of the ground truth. In such a case, unsupervised learning through genetic algorithm with simulated annealing has shown major success in determining sunspot umbra and penumbra~\cite{2018PASP..130j4503Y}. Similarly, self-organizing maps (SOMs) have been employed with great success on spectroscopic data~\cite{2021MNRAS.503.2676S_segment}. These unsupervised neural networks project high-dimensional data onto a low-dimensional grid while preserving the topological structure of the data, so that physically similar profiles map close to each other. Gaussian Mixture Models (GMMs) have shown great success in segmenting not only sunspots, but even the small-scale fine structure in sunspots~\cite{2023A&A...670A.132G}. They model the overall data distribution as a combination of multiple multi-dimensional Gaussian sub-distributions, allowing for soft probabilistic clustering rather than hard boundaries. Classical segmentation and contouring operators have also been tested for sunspot detection~\cite{2025ApJ...980..261C, 2025ApJS..280...70C_ar_detect}, and they are scaled up with deep learning models like U-Net.

While automated methods have been developed specifically for sunspots owing to their fine structure and global impact, such methods have also been developed for other features through straightforward application of {\ml} inference (\infer). Filaments, which we have also seen in Fig.~\ref{fig:aia_features}, are important structures that can become unstable and launch as CMEs into space. Filament identification and study is hence very important to track evolution, understand trigger of instability, and model the evolution of the flux rope after instability. Studies have been dedicated to identifying filaments through spectroscopic data and k-means clustering~\cite{2020A&A...640A..71K}. This unsupervised algorithm partitions data into $k$ distinct groups by iteratively minimizing the distance between data points (e.g., spectral profiles) and their group centroids. The same filaments, when viewed outside of disk, appear bright rather than dark due to no emission in the background. Such regions can also be identified using a U-net formalism~\cite{2024ApJS..272....5Z_prominence}. Fibrils are elongated structures in the chromosphere tracing magnetic field. These can be accurately identified through a U-net formalism~\cite{2021ApJS..256...20J_halphafibrils}. The chromosphere also exhibits vortex motions of plasma, which have been named \emph{chromospheric swirls}. Identifying chromospheric swirls requires mapping the geometrical symmetries of the plasma. By pre-processing curvature-based features, unsupervised clustering operators ($\mathcal{I}$) can successfully isolate these vortex motions~\cite{2021SoPh..296...17D_cluster}. Such feature generation can however be circumvented to the application of symmetry-aware methods and large amount of labeled data.

In the upper atmosphere, two of the most prominent morphological structures are active regions and coronal holes, which we have seen in Fig.~\ref{fig:aia_features}. Coronal hole segmentation is very important to quantify the amount of open flux in the solar atmosphere, which contributes to both the solar wind in the heliosphere and also acts to determine the strength of the next solar cycle. Some methods to estimate coronal holes include using shallow {\ml} supervised models ~\cite{2015JSWSC...5A..23R_segment}, contour based segmentation \& c-means clustering~\cite{2020SoPh..295..110B_coronalhole}, unsupervised k-means clustering~\cite{2022ApJ...930..118I_coronalhole}, and CNN-based supervised models~\cite{2020ApJ...903..115I_coronalhole}. A major failure mode of all of these aforementioned coronal hole identification methods is a confusion between coronal holes and filaments. Filament identification uses passbands sensitive to different temperatures when compared to coronal holes, so as to obtain the most contrast in the features intensity-wise. This confusion is mitigated by using multi-wavelength input data. Additionally, models are trained on clean datasets that explicitly eliminate the coronal hole---filament ambiguity~\cite{2021A&A...652A..13J}.

\subsubsection{Semantic Classification and Feature Representation}
However, more than the identification of sunspots, it is also important to identify the \emph{group of the sunspot}. These are visually assigned labels to sunspots that capture the spatial complexity and structure of the sunspots. This is further related to the probability of instability in these sunspots (for e.g. through solar flares), and the activity of the Sun itself. To address this, cutouts of the identified sunspot are classified into different groups. Classification models include Deep Neural Networks (DNNs)~\cite{2019AdAst2019E..27F}, CNNs~\cite{2024RAA....24i5022M}, and XGBoost~\cite{2021ApJS..257...38T_sunspot}. DNNs are multi-layer feedforward networks that learn abstract physical mappings by passing inputs through sequential layers of non-linear neurons. In contrast, XGBoost is an ensemble method that builds decision trees sequentially to correct the errors of previous trees via gradient boosting. However, efforts have also focused on a full-pipeline of sunspot identification, clustering and classification through the application of multiple neural networks~\cite{2023JGRA..12831548S}. We however note that the full computation graph is not tied in together in such cases, and each \emph{task} is handled independently by an individual network, which does not exchange information with other networks. Recent work on supervised classification of sunspots into groups, however, clearly shows that vision transformers outperform CNNs, and point towards magnetograms being a better source of data than continuum~\cite{2025ApJ...981..157L}.

Apart from single feature extraction methods, there has been significant efforts in a zero-shot identification (\infer) of multiple features using a trained model. For instance, coronal holes, active region core, and active region boundaries can be distinguished through SVD + Bayesian classifier~\cite{2006SoPh..239..519D_segment}. The algorithm considers intensities in different passbands as a feature set, computes a decision boundary on SVD-projected space, and uses these labels for a supervised classification. A more straightforward algorithm is the application of Gaussian mixture model for segmenting our coronal holes and active regions~\cite{2020SpWea..1802478U}. In the lower atmosphere, for instance, cutouts of images, consisting of predominantly one class, are successfully classified by VGG-style networks~\cite{2019SoPh..294...80A_image_classify}. Global segmentation of images based on multi-channel data, based on Faster R-CNN, provides bounding box for different features in a zero-shot manner after training~\cite{2021SoPh..296..160B}. Development of segmentation methods in absence of a ground truth is very tricky. To this end, it is imperative to also evaluate the results from {\ml} models to quantify ``ground truth". For example, ~\cite{2025SoPh..300...89S} develop a global algorithm that develops bounding box for different chromospheric features based on Solar Ultraviolet Imaging Telescope~\citep[SUIT;][]{Tripathi_2017} onboard Aditya-L1~\cite{AL1}. This work trained a YOLO (You Only Look Once) model on synthetic data, modeling the instrument properties of SUIT and IRIS. The model frames target classification and bounding box localization as a single regression task evaluated in a single forward pass. The main innovation is that statistical properties (like entropy and skew) were computed for all features. These were then compared across the synthetic ground truth and the real SUIT observations. This approach helps evaluate model performance when real ground truth is unavailable.

Discrete latent representations offer an alternative to continuous autoencoders for spectral compression. Autoencoders are neural networks that compress data into a low-dimensional latent space and reconstruct it. For example, vector quantization of IRIS spectra into a finite codebook produces state labels that are intrinsically interpretable as physical activity classes and computationally efficient for real-time classification~\cite{2021A&C....3600473I}. Vision-language representation learning extends the state space from visual to semantic: mapping AIA spatial features to natural language captions enables text-query-based solar event discovery and opens the archive to large language model integration~\cite{2024AdSpR..73.3270B}.

\subsubsection{Temporal Tracking}
All the methods presented above consider sunspots data independent of the time coordinate. The underlying assumption is that the training set covers sunspots of almost every possible configuration and property, and hence the {\ml} method is able to identify it correctly. However, an important aspect of feature extraction in solar physics is temporal consistency, where in sunspots actually rotate from east to west, along with an evolution with time. This motion and temporal consistency to track active regions is formulated as a temporal propagation (\proppredict), done for instance by ~\cite{2024SoPh..299..121G_ar_detect}. In this work, the authors develop a feature detection module, a Kalman filter based position prediction, and a temporal association module based on flow graph. This ensures a strong temporal continuity across different frames in the active region.

\subsubsection{Historical Data Rescue}
Apart from current solar data, there is a wealth of archival data covering sunspot drawings across centuries. Digitization of this data via inference operators (\infer) extends the available dataset capability backward in time. Convolutional networks applied to century-long photographic sun-charts extract solar limb, disk center, and plage boundaries from hand-drawn records. This converts analogue archives into ML-ready labeled datasets for solar cycle reconstruction~\cite{2026ApJS..283...19M}. Tabular historical records also present significant digitization challenges due to handwriting variability. For example, the handwritten catalogs from the Zurich Observatory contain coordinates of sunspots, prominences, and faculae that bridge Spörer's early observations with 20th-century records. These were digitized using a convolutional recurrent neural network (CRNN) with an adaptive learning approach. The model automatically retrained on each page to leverage internal dependencies among the coordinate table values~\cite{2022SoPh..297...79I}. 

\subsection{Transient activity}
The coherent structures identified in Section~\ref{subsec:trans_segmentation} exist in unstable equilibria dictated by the interplay between energy injection ($\tau_{\rm in}$) and dissipation ($\tau_{\rm diss}$) timescales. If the time scale of energy injection is shorter than cooling time scales, there are discrete \emph{jumps} of energy in the system. For very slowly varying heating, discrete \emph{jumps} in energy are not as prominent. One more complication is traveling waves --- these are hydrodynamic or MHD waves, which result in oscillations across the whole of the solar atmosphere, and may or may not result in energy deposition. Studying these events is therefore imperative to understand the general stability and energetics of the solar atmosphere.

\subsubsection{Solar flares}
Solar flares provide a primary example of this transient activity. Solar flares are sudden release of energy in the solar atmosphere that manifest as intense, short-lived brightenings across the electromagnetic spectrum. This rapid energy release drives global atmospheric responses, including the launch of fast-mode waves across the solar disk. Furthermore, a solar flare may also result in mass flows, and launching of \emph{filaments} across the interplanetary space, resulting in Coronal Mass Ejections (CMEs). The scale of impact of such events thus prompts detailed statistical studies of both flares, and regions that give rise to flares.

The {\ml} applications in flare physics fall into several distinct operator classes. The first critical task is using a temporal propagation operator (\proppredict) to forecast signatures of instabilities that would be potentially correlated with flares. Static point-in-time boundary conditions are physically insufficient for forecasting without dynamic context, such as flux emergence rates and helicity injection. Because flares are rare, standard loss functions suffer from an ``all-clear'' bias, so training set rebalancing may help learning but the test set must remain untouched to reflect physical reality.

A very well-prepared and standard dataset containing active region data is from features extracted from magnetic field measurements of the photosphere~\citep[SHARP;][]{2014SoPh..289.3549B}. The SHARP dataset has been extensively used to study the underlying physics of active region producing flares, and also to develop {\ml} models to forecast flares. Typically, a given active region can become unstable and flare, but not all active regions flare. To this end, understanding parameters that may be correlated with flaring active regions is imperative. Using the SHARP data, this may be formulated as a classification of active regions into flaring or non-flaring. Such an exercise can be done even with simple models like Support Vector Machines (SVMs), and reveals that the total unsigned vertical current is the most important differentiating parameter between flaring and non-flaring regions~\cite{2015ApJ...798..135B}. Similarly, the mean free magnetic energy and the past flaring history of an active region are also important indicators of flare probability of active regions~\cite{2025FrASS..1109061Y_flarefeatures}. Simple {\ml} models, with extracted physics features, are also able to discover the flare productivity of active regions several days prior to the flare~\cite{2019PNAS..11611141D}.

Apart from the SHARP features, there has been progress in deriving topology-based features to study their impact on flare forecasting algorithms. Such effort shows that topological parameters based on connectedness of sets~\cite{2023A&A...674A.159D} or on the magnetic topology of the active region~\cite{will_2026_topologicalfeatFlares} already contain enough information, and may not require highly complex {\ml} models for such classification. Similarly, k-shape clustering has been used to great effectiveness to identify step-like changes in the horizontal magnetic field during a flare, which is strongly correlated with flare dynamics and activity~\cite{alanazi_2026_BstepdetectFlare}.

Sometimes, the particular SHARP parameters may not be available, since the vector magnetic field data required to compute these parameters may not be at a good enough cadence to capture the time-scale of activity during a flare. Hence, there has been efforts to translate from high cadence AIA intensities~\cite{2021SoPh..296..163B_sharp_from_aia} and line-of-sight magnetic field~\cite{2022ApJ...939...64D}. These efforts act as cross-domain propagation operators ($\mathcal{P}$), translating across states, and correspond to
\begin{equation}
    \mathcal{P}_{\rm data \rightarrow SHARP} : \vec{S}_{\rm data} \rightarrow \vec{S}_{\rm SHARP}.
\end{equation}

While forecasting models attempt to predict flares, inference operators (\infer) diagnose them in real-time. While the photosphere shows indications of system stability, the atmosphere is known to already start responding to energy accumulation before the start of a flare. For instance, spectral lines in the chromosphere already show indications of flaring activity several minutes before a flare~\cite{2018ApJ...861...62P, 2021ApJ...922..137W, 2021A&C....3600473I,2024A&A...689A..72Z,2026ApJ..1001..104Z}. The key methodological insight is that clustering is not merely dimensionality reduction, but physical taxonomy. When hierarchical k-means is applied to \ion{Mg}{ii}~h\&k profiles during flaring conditions, the resulting clusters directly map to distinct atmospheric regimes --- quiet-Sun double peaks, pre-flare single-peaked heated profiles, and broad blueshifted-reversal profiles at ribbon fronts --- and the last cluster correlates directly with hard X-ray signatures of high-energy electron deposition~\cite{2018ApJ...861...62P}. This connection between spectral morphology and particle acceleration physics makes the clustering result a physical discovery, not just a compression.

Inherently, the shapes of spectral lines that form in optically thick conditions depend on the atmospheric column properties, and these shapes can be multi-peaked non-Gaussian profiles. Hence, a systematic study of different spectral lines and their inter-correlations during flares can be expressed through a formalism of mutual information. A very innovative way of performing this would be to transform continuous spectra to categorical data using k-means clustering, and then compute the mutual information across different spectral lines~\cite{2021ApJ...912..121P}. A very interesting paradigm is to rephrase the solution through a mutual information neural estimator to study such spectral associations. The complexity of spectral observations, reduced by {\ml}, reveals the inherent coupling across spectral lines which occurs due to inherent coupling across the atmospheric column. Thus, application of both information theory and {\ml} enables interpretation of potentially even new physics from complex spectral observations~\cite{2021ApJ...912..121P}.

Solar flares are typically inferred from soft X-ray observations. However, if the flare footpoint is behind the solar disc, a visual inspection is needed to decipher if the spectral response corresponds to a flare or not. However, parameterizing the problem as classification using spectrograms enables identification of occulted flares~\cite{2021SoPh..296...39I_occulted_flares}. 

A single energetic flare can trigger a cascade of secondary atmospheric phenomena that ripple across the entire solar disk. Following solar flares, the whole atmosphere may respond to release of energy, through the propagation of fast-mode waves across the atmosphere, known as EUV / EIT waves. Detecting these waves enables us to determine the energetics and properties of the solar flare, and the overarching environment in which the active region rests. Furthermore, with CMEs, they are also able to provide early indicators of onset of CMEs, enabling space weather preparedness of humanity. The particular \emph{cutout} containing the wavefront can be identified by models like Faster Region-based CNN (Faster R-CNN), and the outline of the wavefront drawn using a pix2pix GAN formalism~\cite{2020SoPh..295...44X_euv_wave}, enabling a very fast identification of wavefronts due to flares.

\subsubsection{Other transient events}
While solar flares are excellent examples of transient events in the solar atmosphere, they are the largest of such events, in terms of energy, length and time scales. The transient events occur at a wide variety of time scales, occurring across the solar atmosphere. Since the amount of energy in such events might be small, the atmospheric response may not be observed across all spectral lines / wavelength bands. 

In the photosphere, concentration of magnetic flux elements are seen as bright points. Identifying, tracking and studying these bright points presents us with an understanding of magnetic field dynamics in the photosphere, which are used to understand the energy injection into the solar atmosphere. To isolate these transient bright points from the photospheric background, {\ml} inference operators (\infer) estimate the exact separation thresholds directly from the observational data~\cite{2023ApJ...956...62B_brightpoints}.

In the chromosphere, Ellerman bombs are very compact ($\sim1000$ km), dynamic (lasting $\sim5$ min) events seen in lower atmosphere. CNN and DNN models, trained on both high resolution spectra and low resolution imaging data are able to identify these events very accurately~\cite{2025A&A...699A..54S}. Such methods are also validated by studying the variation of performance through permutation of inputs, and show consistency with known physics. Apart from Ellerman bombs, the chromosphere also displays a myriad zoo of features, including rapid flows seen as elongated streaks in space, known as spicules --- chromospheric plasma jets that are thought to be a key channel for mass and energy transport into the solar corona. k-means clustering has proven to be extremely successful in identifying spectral profiles exhibiting different kinds of spicules~\cite{2019A&A...631L...5B_rbe_rre_kmeans}. 

Much higher in the transition region and corona, small-scale impulsive events are observed again. Parker's nanoflare hypothesis proposes that the corona is sustained by a superposition of sub-resolution reconnection events --- individually too faint to detect, but collectively sufficient to maintain coronal temperatures above $10^6$~K. For instance, combining Zernike moments of images with an SVM classifier works effectively in scaling up identification of small-scale events with the SolO/EUI~\cite{2022A&A...663A.128A_smallscaleeuv}. Sometimes, the individual small-scale impulsive events may not be directly observable. Rather, what may be observed is the impact of the statistical distribution of such events. To estimate the distributional parameters of these events, one may perform a simulation based inference by training on a set of synthetic data, and performing inference on real data~\cite{2008A&A...492L..13B_nanoflare,2021ApJ...916...59U_nanoflare,2022ApJ...940L..38U_nanoflare}. These methods push the limits the of the instruments, and enable large scale statistical studies of the smallest transients in the solar atmosphere. Such simulation based inference can also be performed through a realistic physics simulator. By training a model on features extracted from simulations, a Decision Tree or Random Forest can provide insights into the typical heating mechanism in the active regions~\cite{2021ApJ...919..132B_arcores_heating}. A Random Forest makes predictions by aggregating the consensus vote of many individual decision trees trained on different data subsets to avoid overfitting. Such weak impulsive events are also observed in radio imaging spectroscopy. They are identifiable using a combination of t-distributed Stochastic Neighbor Embedding (t-SNE) and Density-Based Spatial Clustering of Applications with Noise (DBSCAN) algorithms~\cite{2023ApJ...954...39B_smallscale_id}. Here, t-SNE acts as a non-linear dimensionality reduction technique that maps high-dimensional embeddings to 2D or 3D spaces for visual diagnostics.

Radio burst observations present a disk-integrated spectral response of the solar atmosphere. These signatures arise due to deceleration of accelerated electrons, and have a peculiar radio spectrum. Identifying and classifying these bursts is traditionally manual, making it an ideal target for automated inference operators (\infer). Using You Only Look Once (YOLO)-based~\cite{2024A&A...683A..90Z_radioburst} and GAN-based models~\cite{2021FrP.....9..428Y_radioburst}, researchers have demonstrated that {\ml} inference models (\infer) can successfully identify these signatures in a rather straightforward manner. Specifically, given a full spectral observation,~\cite{2024A&A...683A..90Z_radioburst} demonstrate bounding boxes around the radio bursts, including the particular type of radio burst. Such automation processes enable a statistical study of radio and EUV signatures of events occurring in the heliosphere.

In a similar vein, reconnection events which possibly give rise to the non-thermal electrons causing radio bursts can be simulated using kinetic plasma simulations. Unsupervised methods like Bayesian Gaussian Mixture models model electron velocity distribution as a sum of thermal and non-thermal components, effectively identifying non-thermal electrons in simulations~\cite{2022PhPl...29b2104Y_reconnection_simulation}. Such physics-informed modeling methods make the model transparent, and present accurate constraints on the physics.

Apart from such drastic transient events, travelling waves are very common in the solar atmosphere. Recurrent networks applied to IRIS Doppler time-series characterize wave physics in the transition region by identifying $25$--$40$~second damping lags between chromospheric and transition region bright point oscillations --- demonstrating that learned temporal sequence analysis can extract physically meaningful propagation timescales~\cite{2026SoSyR..60...11Y}. Similarly, CNN-LSTM models can differentiate between images with/without kink oscillations in the solar corona~\cite{2025ApJS..281...12B_kink_oscillation}. In such methods, a t-SNE visualization of the LSTM embeddings clearly differentiates between the two classes naturally, providing support in favor of model \emph{learning} the physics in some sense.

While clustering strategies map standard atmospheric regimes into discrete physical states, they are fundamentally blind to extreme physical anomalies and rare transient phenomena—such as intense localized heating, high-velocity jets, or small-scale reconnection events—which are often lost when {\ml} models optimize for average statistical trends. To isolate these highly informative physical extremes, unsupervised density estimators can be deployed. For example, Inspectorch~\cite{2026arXiv260220316D} operates across heterogeneous solar data archives (including Hinode/SP, IRIS, SST/CHROMIS, SDO/AIA, and Solar Orbiter/EUI) to assign probabilistic anomaly scores directly to observed profiles. Rather than relying on manual catalogs, this framework isolates physically significant outliers—such as chromospheric profiles with uncommon broadening, extreme Doppler shifts, or transient EUI coronal brightenings—offering a systematic way to filter massive datasets and focus computational analysis on the most energetic and rare physical processes.
\section{Observation-space operators: instrument correction, cross-mission harmonization, and instrument design}
\label{sec:cross_instrument}

This family of applications cuts across traditional physical layers. It operates instead as the mathematical link that harmonizes heterogeneous data archives, synthesizes virtual instruments, and learns unified representation spaces. Rather than modeling temporal evolution, these models construct learned mapping operators (\propcouple) that couple disparate instruments, viewpoints, wavelengths, historical archives, or even semantic descriptions. 
\begin{svgraybox}
Mathematically, these applications are formulated through coupling operators that act on the solar state vector or translate between distinct observational representations:
\begin{equation}
\label{eq:translation_coupling}
\mathcal{P}_{a\rightarrow b} : \vec{S}_a \rightarrow \vec{S}_b \quad \text{or} \quad \mathcal{P}_{1\rightarrow 2} : \vec{O}_1 = \mathcal{O}_1[\vec{S}] \rightarrow \hat{\vec{O}}_2 = \mathcal{O}_2[\vec{S}],
\end{equation}
where the source state $\vec{S}_a$ and the target state $\vec{S}_b$ may represent observational encodings (e.g., detector pixel arrays, spectral profiles) or latent physical parameters (e.g., localized vector magnetic fields, plasma temperatures).
\end{svgraybox}
To systematically review this diverse literature, we categorize these methodologies into three distinct, interconnected thematic families defined by the mathematical nature of their mapping operations, their role within our state-vector taxonomy, and the components of the solar state vector they act upon. The first family focuses on correcting local atmospheric and optical degradations within a single instrument setup. These methods solve classical inverse problems to reverse localized physical distortions---like atmospheric turbulence, point spread function broadening, or polarimetric noise---and recover the unperturbed physical state. Crucially, this family also includes active atmospheric correction via adaptive optics. This represents the first active deployment of the control (\control) operator in solar instrumentation, using reinforcement learning to calculate optimal deformable mirror commands in real time. The second family covers cross-mission harmonization for multi-generation studies. Rather than inverting local physical degradations, these networks act as domain-to-domain translation operators ($\mathcal{P}_{1\to 2}$). They map observations across distinct hardware generations (e.g., SOHO/MDI to SDO/HMI) to correct for differences in pixel grids and detector aging, establishing consistent solar climate records. The final family exploits the strong vertical thermodynamic coupling and magnetic connectivity in the solar atmosphere. These models execute cross-wavelength image translation and virtual instrument synthesis. Serving as cross-domain coupling operators (\propcouple), they translate observations across height-dependent passbands to synthesize missing channels or predict integrated EUV irradiances.

No single instrument can observe the full solar state, making cross-representation translation crucial for placing separately observed subsystems into a common, unified state-space framework. However, a major physical challenge in this domain is the risk of ``hallucination"—generative models introducing non-physical details to satisfy mathematical loss profiles. To address this, state-of-the-art architectures increasingly incorporate physics-informed loss constraints (such as divergence-free magnetic bounds or radiative transfer consistency). They also employ rigorous cross-validation against held-out space missions. Finally, downstream scientific validations are performed to ensure that translated states represent physically plausible solar phenomena. Furthermore, standard convolutional operators applied to full-disk images inherently assume a flat grid, ignoring projection effects and spherical geometry unless explicitly corrected.

\subsection{Correcting local atmospheric and optical degradations}
\label{subsec:trans_deconvolution}

Operating strictly within the domain of a single instrument setup, the first family of applications is formulated around reversing the localized optical distortions, atmospheric aberrations, or detector noise that degrade the measured solar state. This is depicted through modeling $\mathcal{O}_{\mathrm{instr}}$ or $\mathcal{O}_{\mathrm{instr}}^{-1}$ in Fig.~\ref{fig:forward_op}. 

The physical state vector $\vec{S}_{\rm dec}$ represents the unperturbed spatial or polarimetric scene, the wavefront phase distortions, or the mechanical deformable mirror states:
\begin{eqnarray}
\vec{S}_{\rm dec}: \mathcal{O}_{phys}[\vec{S}] = \{I, Q, U, V, \phi, \vec{a}_{\rm AO}, \ldots\}(x,y,\lambda,t),
\end{eqnarray}
where $I$, $Q$, $U$, and $V$ represent the four Stokes profiles containing the full polarimetric signature of the solar plasma; $\phi$ is the atmospheric wavefront phase aberration representing optical phase delays; and $\vec{a}_{\rm AO}$ represents the mechanical actuator correction commands of an active adaptive optics system. Mathematically, the spatial degradation process can be formulated as a non-linear forward convolution of the true physical scene with a spatially varying Point Spread Function (PSF), added with polarimetric or detector noise:
\begin{equation}
\label{eq:deconvolution_forward}
\vec{O}(x,y) = \iint \vec{S}_{\rm dec}(x',y') \cdot \text{PSF}(x-x', y-y'; x, y) \, dx' dy' + \vec{\eta}(x,y),
\end{equation}
where $\vec{O}$ represents the degraded multi-channel observations and $\vec{\eta}$ represents stochastic noise. In this context, {\ml} acts as an inverse operator (\infer) to reconstruct the unperturbed physical state $f_\theta(\vec{O}) \approx \vec{S}_{\rm dec}$, an emulator (\emulate) to represent the convolving forward process, or a controller (\control) to calculate deformable mirror actuator corrections in real time.

\subsubsection{Denoising}
Recovering sub-resolution solar structure from seeing-degraded observations is an ill-posed inverse problem. The convolving PSF is spatially variant, partially unknown, and entangled with stochastic noise, making analytical inversion completely unstable without strong regularization. To address detector and photon noise, self-supervised and generative architectures have been deployed to bypass simulated ground-truth reference data. The first self-supervised approach—structurally analogous to noise-to-noise training in machine vision—exploits the spatial and spectral coherence latent in the observations themselves. By using neighboring pixels and wavelength channels as implicit mutual constraints, these models recover weak polarimetric signals at or near the detector noise floor from real observations without requiring simulated training pairs~\cite{2019A&A...629A..99D}. Alternatively, generative models trained on paired short- and long-exposure data replicate temporal averaging from a single noisy frame, effectively reconstructing the photon statistics of a much longer integration without sacrificing temporal cadence~\cite{2020ApJ...891L...4P}.

\subsubsection{Spatial PSF deconvolution}
To address spatial seeing aberrations, a deeper transition shifts machine learning from static supervised regression to physics-guided inversion of analytical image formation models. The key methodological insight is \emph{differentiable physics integration}: by embedding classical optical diffraction theory directly into the neural loss function, the wavefront aberration becomes a latent variable optimized jointly with the scene, so the PSF need not be known in advance. This blind deconvolution approach parameterizes both the convolving wavefront and the unperturbed scene as latent variables optimized against bursts of observed frames, accelerating classical reconstructions by three orders of magnitude~\cite{2021A&A...646A.100A}. 

Anisoplanatism—the spatially varying turbulence that makes the PSF different at every point in a wide-field image—breaks the single-PSF assumption and produces stitching artifacts when handled patch-wise. The resolution is a continuous coordinate-based forward emulator (\emulate) for wide-field PSF convolution that eliminates patch boundaries~\cite{2024A&A...688A..88A}. This has been extended to implicit neural representations. These models simultaneously optimize continuous photospheric scenes and anisoplanatic PSFs directly on ground-based data~\cite{2026A&A...708A.170S} (see also Fig.~\ref{fig:forward_op}). Furthermore, temporal coherence constraints enable attention-based fusion of sequential observations: explicit inter-frame alignment using deep convolutional architectures suppresses residual wavefront noise while preserving fine-scale photospheric and chromospheric structures~\cite{2021A&A...652A..50W,2023A&A...674A.126Z}. For traditional supervised approaches, simulation-based inference remains a powerful tool: radiative-MHD simulations (e.g., MURaM) provide sub-resolution Stokes profiles as ground truth, enabling convolutional inverse operators ($\mathcal{I}$) trained on synthetic pairs and validated against high-resolution Hinode observations~\cite{2018A&A...614A...5D}.

\subsubsection{Spectral PSF deconvolution}
Spectral and polarimetric degradations constitute a distinct class of instrument-space distortions, separable from spatial PSF broadening and corrupting different components of the state vector. Line-profile smearing in spectrographs erases Doppler velocity and thermal line-width information, whereas non-linear saturation in filter-based magnetograph Stokes signals suppresses inferred field strengths. Spectral PSF deconvolution is the most direct solution: every spectrograph broadens line profiles through a known instrumental response, and learned convolutional surrogates trained on this forward model successfully recover the true physical line shape at science cadence, bypassing the computational cost of iterative deconvolution~\cite{2020FrASS...7...29M}.

\subsubsection{Wavefront sensing and adaptive optics control}
Active compensation of atmospheric turbulence transitions {\ml} from passive post-facto state estimation to real-time physical control. Classical adaptive optics (AO) reconstructors are linear: they invert a static interaction matrix measured during calibration to map wavefront sensor slopes to actuator commands. This linearity fails under the conditions most demanding for solar observations—extreme photon starvation on faint targets, non-stationary turbulence at daytime ground-based telescopes, and actuator cross-coupling not captured by the calibration matrix. 

Transcending these limits requires a policy that can learn the non-linear actuator-to-wavefront dynamics online and adapt to changing conditions without re-calibration. ~\cite{2022A&A...664A..71N} established this with the Policy Optimization for Adaptive Optics (PO4AO) framework. This was the first deployment of the \control\ operator in solar instrumentation. The model-based reinforcement learning policy $\pi_\theta(\vec{a}_{\rm AO} | \vec{s}_{\rm WFS})$ maps pyramid wavefront sensor states directly to deformable mirror actuator commands. It learns this mapping by maximizing the Strehl ratio reward $R(\vec{s}_{\rm WFS}, \vec{a}_{\rm AO})$. Non-common-path aberrations (NCPA) represent a second class of residual that linear reconstructors cannot address by design. These are static and dynamic optical errors downstream of the wavefront sensor and are therefore invisible to it. ~\cite{2026arXiv260400993N} solved this with PO4NCPA, using sequential phase diversity to execute active focal-plane control without prior calibration. A third physical limit is phase sign ambiguity: a single intensity image is symmetric under phase conjugation, so sign-correct wavefront recovery from a single frame is underdetermined. ~\cite{2025OExpr..33.9224Z} constrained this through multispectral phase-shift measurements, while ~\cite{2026OExpr..34.7122Z} resolved it explicitly via the Physically Optimized Single-Image Sensorless (POSS) method, embedding physics-guided sign constraints into a lightweight Residual Network (ResNet-18) to achieve millisecond phase diversity accuracy.

\subsubsection{Saturation artifact correction}
During high-energy flaring events, extreme EUV and X-ray emission frequently saturates the CCD pixels of space-based imagers, leading to severe information loss in the flaring cores. Rather than relying on simple linear interpolation or discarding the data, deep learning models exploit spatial correlations from surrounding, non-saturated pixels to reconstruct peak flare intensities at full spatial and temporal resolution~\cite{2022RAA....22f5009Y}. This enables accurate thermodynamic and magnetic diagnostics even within the most extreme, energetic regimes of solar active regions.

Deconvolution, denoising, and wavefront control act as structured inverse operators (\infer), emulators (\emulate) or controllers (\control) that restore the unperturbed solar state vector under optical degradation. However, their physical accuracy remains bounded by polarimetric noise floors and calibration limits. No super-resolution network can transcend these hardware limits. Visual sharpness in super-resolved magnetograms is physically insufficient; the generative operator (\propcouple) must strictly preserve magnetic flux conservation to avoid hallucinating unphysical fields. The deepest open problem is not spatial fidelity but physical uniqueness: guaranteeing stable, sign-correct wavefront solutions under anisoplanatic wide-field turbulence and photon starvation is a constraint that current reinforcement learning controllers do not yet satisfy.

\subsection{Temporal consolidation and archive homogenization}
\label{subsec:trans_calibration}

While deconvolution and denoising operate within a single instrument to remove localized optical distortions, the second family maps observed solar states across distinct hardware platforms, space missions, or generational baselines. This domain is defined by the homogenized physical state vector $\vec{S}_{\rm cal}$:
\begin{eqnarray}
\vec{S}_{\rm cal} = \{B_{\rm LOS}, \vec{B}_{\rm phot}, I_{\rm cont}, \ldots\}_{\rm hom}(x,y,t),
\end{eqnarray}
where $B_{\rm LOS}$ represents the line-of-sight magnetic field, $\vec{B}_{\rm phot}$ the vector magnetic field, $I_{\rm cont}$ the continuum intensity, and the subscript ``${\rm hom}$'' denotes a standardized spatial scale, detector sensitivity, and calibration baseline. Because these networks map a source instrument's observational space to a target instrument's standard without a physical forward operator, they are formulated as cross-domain coupling operators $\mathcal{P}_{1\to 2}$:
\begin{eqnarray}
\mathcal{P}_{1\to 2} : \vec{O}_1 = \mathcal{O}_1[\vec{S}_{\rm cal}] \rightarrow \hat{\vec{O}}_2 = \mathcal{O}_2[\vec{S}_{\rm cal}],
\end{eqnarray}
correcting for generational differences in pixel grids, spectral responses, and sensor aging to establish consistent, multi-decade solar climate archives.

\subsubsection{Multi-mission standardization and domain homogenization}
Solar cycle studies, irradiance reconstructions, and space climate models require magnetic and intensity baselines that span multiple instrument generations—a requirement that no single mission satisfies. The observational gaps, spatial resolution jumps, and calibration discontinuities between successive missions introduce systematic offsets that can masquerade as real solar variability if left uncorrected. The core methodological difficulty is supervision: co-temporal image pairs from two instruments are often unavailable because the instruments operated in different eras or at incompatible cadences. 

Unpaired domain translation resolves this by learning mappings from unpaired datasets rather than paired examples. It uses adversarial training with Cycle-Consistent Generative Adversarial Networks (CycleGANs). These networks learn to translate images between two domains without co-temporal training pairs. They enforce cycle consistency: translating an image from domain A to B and back to A must reconstruct the original image~\cite{2025NatCo..16.3157J}.

The SOHO/MDI--to--SDO/HMI transition is the most consequential instrument gap in the modern magnetogram record, separating 16 years of continuous monitoring from the high-cadence, high-resolution HMI era. Bridging it requires resolving not only spatial resolution differences but also systematic field-strength biases, detector calibration offsets, and non-stationary drifts caused by sensor aging and orbital variations over the mission lifetimes. The key advance over generic super-resolution is embedding known physical degradation kernels into the learned inverse mapping. By including the measured PSF, pixel scale, and calibration curve, the network corrects for physically characterized distortions rather than inferring them from data alone. Validation against scientific benchmarks (such as flux distributions and active region complexity) rather than visual metrics then tests whether the recovered states support downstream physical analysis~\cite{2024ApJS..271...46M}. Temporal gaps require a different strategy. Denoising Diffusion Probabilistic Models (diffusion models) generate data by learning to reverse a gradual noise-injection process, preserving sharp high-frequency details and statistical textures. These models are deployed to interpolate sequences across gaps, preserving dynamic flux evolution rather than blending adjacent frames~\cite{2025ApJ...980..228L}. Additionally, neural networks can estimate absolute magnetic flux directly from continuum intensity images, serving as a valuable tool for observational historical flux reconstructions~\cite{2021A&A...652A.143B_Bfromcontinuum}.

Spatial super-resolution is a critical sub-component of this homogenization process. Rather than simply applying traditional sharpening filters, deep learning models estimate unresolved magnetic structure by utilizing priors learned from higher-resolution instruments or synthetic simulations. Following early baseline enhancements of SDO/HMI~\cite{2018A&A...614A...5D}, subsequent architectures have extended this capability to both line-of-sight and vector magnetograms across HMI, MDI, GST/NIRIS, and related data products~\cite{2020ApJ...897L..32R, 2024SoPh..299...36X, 2024RAA....24f5029Q, 2025A&A...697A.110X}. The practical scientific question governing these operators ($\mathcal{P}_{1 \to 2}$) is whether the recovered small-scale flux, gradients, and polarity-inversion lines are physically reliable enough to support derived calculations, such as helicity proxies or flare-prediction features.

The data-rescue problem of handling centuries of heterogeneous historical conventions is fundamentally distinct from the high-throughput processing required for modern, highly cadenced observatories. Before the era of systematic magnetograph surveys, the only available proxies for surface magnetic field were chromospheric diagnostics—H$\alpha$ plage morphology, \ion{Ca}{ii}~K intensity—which encode magnetic information indirectly through the plasma response to concentrated flux. Plage regions are areas of enhanced chromospheric brightness co-spatial with concentrated photospheric magnetic flux; their historical brightness in H$\alpha$ and \ion{Ca}{ii}~K filtergrams thus serves as the primary magnetic field proxy for the pre-magnetograph era. Reconstructing a pseudo-magnetogram from these proxies is a physical wager: it is meaningful only if the magnetic-to-chromospheric coupling is sufficiently regular to invert statistically, an assumption that depends on activity level and spatial scale. Conditional GANs trained chronologically on co-temporal H$\alpha$ and HMI pairs can synthesize pseudo-magnetograms while preserving magnetic polarity consistency, extending quantitative magnetic field estimates into the ground-based photographic archive~\cite{2023ApJS..266...19G}. The same chromospheric-to-magnetic inversion holds for \ion{Ca}{ii}~K filtergrams, where it demonstrates that historical archives spanning decades before the HMI era can serve as quantitative proxies for magnetic field evolution across past solar cycles—provided the chromospheric coupling remains within the regime the training data sampled~\cite{2020ApJ...895L..16S}.

Cross-instrument calibration and domain homogenization act as coupling operators ($\mathcal{P}_{1\to 2}$) whose primary contribution is the reconstruction of standardized multi-decade solar climate records across sensor generations. Two fundamental challenges remain open: spatial super-resolution of magnetograms offers no guarantee of magnetic flux conservation, and unpaired cross-sensor domain mappings risk introducing polarity inconsistencies that propagate undetected into long-term solar cycle analyses.

\subsection{Virtual instruments}
\label{subsec:trans_wavelength}

The third family of applications exploits the vertical thermodynamic coupling and magnetic connectivity that structures the solar atmosphere from the photosphere to the corona. Rather than mapping the same physical state across different instruments, these models map the solar state across entirely different wavelengths, heights, or spectral profiles. The state vector $\vec{S}_{\rm wave}$ represents this multi-wavelength, stratified atmospheric profile:
\begin{eqnarray}
\vec{S}_{\rm wave} = \{\vec{B}_{\rm phot}, I_{\rm CaIIK}, I_{\rm H\alpha}, I_{\rm EUV}(\lambda_i), F_{\rm EUV}, \ldots\}(x,y,t),
\end{eqnarray}
where $\vec{B}_{\rm phot}$ represents the photospheric vector magnetic field, $I_{\rm CaIIK}$ and $I_{\rm H\alpha}$ the chromospheric \ion{Ca}{ii}~K and H$\alpha$ line intensities, $I_{\rm EUV}(\lambda_i)$ the spatial extreme ultraviolet (EUV) coronal maps at wavelength $\lambda_i$, and $F_{\rm EUV}$ the integrated full-disk spectral irradiance. These applications are formulated as cross-domain coupling operators $\mathcal{P}_{1\to 2}$ that synthesize unobserved channels, or as inference operators $\mathcal{I}_{\rm wave}$ that map spatial features to integrated spectral irradiances:
\begin{eqnarray}
\mathcal{I}_{\rm wave} : \vec{O}_{\rm AIA} = \mathcal{O}_{\rm AIA}[\vec{S}_{\rm wave}] \rightarrow F_{\rm EUV}(t),
\end{eqnarray}
creating ``virtual instruments'' that substitute for absent, damaged, or saturated hardware.

\subsubsection{Spectral channel proxy prediction}
The foundational physical premise of this family is that the solar atmosphere is a vertically stratified, thermodynamically coupled system: the photospheric magnetic field governs the coronal plasma topology, and the radiative coupling between heights means that observations at one wavelength contain latent information about other atmospheric layers. By exploiting these physical correlations as learned cross-domain mappings ($\mathcal{P}_{1\to 2}$), generative models can synthesize coronal or chromospheric diagnostics from photospheric inputs without physical instruments. To achieve this, the generic encoder-aggregator-forecast pattern provides a highly reusable architectural template. Individual encoders extract features from diverse, multi-source solar data. An aggregator then unifies them into a single latent state, and a forecast head maps them to the target observation. For instance, models can cross-calibrate and re-interpret SDO/HMI magnetic field data by mapping it to Hinode/SOT equivalents, lowering the noise floor and resolving disambiguation issues~\cite{2022ApJS..259...24H_synthia, 2024ApJ...970..168W_supersynthia}. However, as demonstrated by~\cite{2025ApJ...993...58F_vectorB_from_los}, these inter-instrument mappings are highly brittle; minute spatial co-registration or alignment discrepancies between the source instruments will severely corrupt the target vector field.

The first demonstration that this coupling is statistically learnable established that conditional GANs and L1-loss optimized CNNs can map line-of-sight HMI magnetic boundary conditions to multi-wavelength AIA UV and EUV images, with adversarial loss preserving the fine spatial contrast of coronal loop structures that pixel-wise losses blur away~\cite{2019ApJ...884L..23P}. A systematic quantification of the physical limits of this learned coupling reveals that spatial morphologies and thermal brightness covariances are faithfully recovered over three orders of magnitude under quiet and weakly active conditions.

The adversarial loss that makes GAN-synthesized images visually sharp is simultaneously the mechanism that can hallucinate physically inconsistent coronal structures: the discriminator penalizes lack of realism, not violation of radiative transfer or magnetic consistency. This architectural tension—between perceptual quality and physical validity—is the organizing design question of the field. Resolving it in favor of physical validity often means replacing adversarial discriminators with adaptive instance modulation architectures. These eliminate spurious high-frequency features at the cost of some spatial sharpness~\cite{2025A&A...702A..83S}. Historical reconstructions push these mappings backward in time: pseudo-EUV ($304$~\AA) images synthesized from HMI magnetograms extend access to chromospheric dynamics into the pre-SDO archive~\cite{2021arXiv210603814D}, while H$\alpha$-to-\ion{He}{i} $10830$~\AA\ translation extends chromospheric diagnostics across more than a century of ground-based records~\cite{2025ApJ...984...99M}.

\subsubsection{Solar irradiance prediction}
When hardware spectrometers are absent, damaged, or suffer severe degradation, predicting the integrated extreme ultraviolet (EUV) solar spectral irradiance from spatial imager observations becomes operationally critical. AIA narrowband coronal images contain sufficient information to reconstruct the full EUV solar spectral irradiance with relative errors below $4\%$, completely bypassing expensive Differential Emission Measure (DEM) inversions~\cite{2019SciA....5.6548S}. This principle has been successfully deployed to compensate for the SDO/EVE/MEGS-A hardware failure, with earlier historical precedents predicting SDO/EVE EUV irradiances using active region magnetic filling factors and radial basis function neural networks~\cite{2013SoPh..283..143C}. Extending irradiance reconstructions further into the pre-SDO era, Bayesian frameworks (\infer) applied to low-fidelity, coarse chromospheric proxies (such as \ion{Ca}{ii}~K and SOHO/SEM) successfully reconstruct full-disk coronal EUV emissions with rigorous, physics-based uncertainty bounds~\cite{2025ApJS..280...50J}.

Cross-wavelength translation and virtual instrument synthesis rest on the physical wager that the thermodynamic coupling between atmospheric layers is regular enough for a statistical model to invert—a wager that demonstrably fails during energetic flares, when the atmospheric state departs furthest from the training manifold. Whether spatial imagers, which project the full emission measure distribution onto a handful of broadband channels, carry the causal information needed to uniquely substitute for a physical spectrometer remains an unresolved foundational question.

\subsection{Instrument design}
\label{subsec:instrument_design}
\begin{svgraybox}
A unifying physical insight underlies this group of applications: the solar atmospheric state, despite being sampled across dozens of spectral channels or hundreds of wavelength points, occupies a manifold of much lower intrinsic dimensionality than the raw observational space. Because the radiative transfer processes that populate each passband are strongly coupled through a shared set of thermodynamic and magnetic parameters—temperature, density, velocity, and field strength as functions of height—the multi-channel data cube is inherently redundant. The full observational vector $\vec{O}$ can often be well approximated by a projection onto a compact latent representation $\vec{z}$ with $\dim(\vec{z}) \ll \dim(\vec{O})$. {\ml} exploits this structure along three complementary axes: \emph{spectral compression}, replacing costly full-profile diagnostics with compact latent encodings; \emph{virtual instrument synthesis}, reconstructing missing or degraded channels from the remainder; and \emph{observational design}, selecting non-redundant wavelengths that maximize physical information per collected photon.
\end{svgraybox}
This redundancy has been quantified empirically across the AIA channel set: training models to predict each passband from all possible subsets of the others reveals that three channels—$131$, $1600$, and $304$~\AA—span the thermal vertical structure well enough to recover the full set under quiet conditions. However, cross-passband translations systematically fail during massive flares because the plasma is driven far outside the thermodynamic manifold sampled during quiet-Sun training~\cite{2021ApJ...915L..31L,2022ApJ...937..100S}. The same compressibility holds in high-resolution line spectroscopy: a $4$-dimensional autoencoder latent space faithfully represents the $110$-dimensional \ion{Mg}{ii}~k profile space sampled by IRIS, with a compression factor of $\sim 27$ and sub-noise residuals, where the latent axes correlate directly with physical diagnostics such as line widths and asymmetries~\cite{2021arXiv210307373S}. Instrument compression models impose specific physical constraints on the latent manifold, distinguishing standard autoencoders from strictly invertible normalizing flows. 

Beyond compressing existing data, understanding this intrinsic dimensionality provides a principled basis for optimizing future observational design. Quantitative analyses of how spectral resolution affects recoverable information content in solar Stokes spectra allow for the deliberate optimization of observing strategies and ML input grids~\cite{2025A&A...693A.272D_spectralres}. The most consequential application of this approach is the direct, non-uniform optimization of wavelength sampling. Rather than distributing exposure time uniformly across a spectral range, neural networks can select optimized wavelength grids for Fabry-Pérot interferometers that maximize Fisher information regarding physical atmospheric parameters, thereby significantly improving spectro-polarimetric inversions without requiring additional photons~\cite{2023A&A...673A..35D}.
\section{Conclusion and future directions}
\label{sec:conclusion}

In this review, we have redefined the Sun through the lens of a state vector characterized by partial observability. This is motivated by the presence of phenomena varying at a range of scales, while at the same time interacting with each other.  Through the operator framework ($\mathcal{I}, \mathcal{P}, \mathcal{E}, \mathcal{C}$), we can compare methods across domains while preserving physical meaning. Such a definition emphasizes how {\ml} serves not merely as an algorithmic optimization, but as a tool for formal state estimation, uncertainty quantification, and dynamical modeling. By mapping the literature to this taxonomy, several synthesis conclusions emerge regarding the current limits and successes of data-driven heliophysics:
\begin{itemize}
\item {\bf The solar interior}: Inference operators for helioseismic and simulation-backed inversions now recover subsurface flow and latent magnetic information at scales that were previously computationally prohibitive. However, the dominant limitation is no longer network capacity, but the domain mismatch between synthetic training environments and real observations. Furthermore, for solar-cycle and dynamo applications, pure sequence learning remains sample-limited. The most credible path is hybrid data-assimilation and physics-constrained forecasting rather than purely black-box extrapolation.
\item {\bf The solar atmosphere}: Radiative transfer, especially in the optically thick regime, is a computationally expensive calculation. Hence, it experiences bottlenecks in both the forward and the inverse operation. Here, {\ml} has proved important as a physics surrogate engine, and as an inverse approximation, including uncertainty estimation. Following from radiative transfer, {\ml} has also been successfully used to generate 3-D representation of the magnetic field, approximating the forward operator, and generating accurate solutions at very high time cadence. We note that most of the applications of {\ml} are generally tied to very specific data, and the application has been very neatly separated out into different radiation regimes, and between the magnetism - radiation domains. A comprehensive understanding necessitates development of methods which are learning a general operator, and not extremely tailored to specific data, which would need to be retrained for different data modality / instrument. 
\item {\bf Coherent structures and transients}: {\ml} has been a powerful tool in detecting structures and transients across a range of length, time and energy scales. The identification strategies have moved past simple detection to physically structured characterization. This enables development of statistical studies of these structures, presenting power constraints on the mass and energy transport across the solar atmosphere. The physics understanding from these {\ml} applications has resulted in the field moving from a rather discrete set of {\it labels} towards a more continuous paradigm. The transition toward unsupervised density estimators provides a systematic way to filter massive datasets without losing the highly informative, rare physical extremes that drive solar transients.
\item {\bf The observation space}: {\ml} has been used extensively in standard data reduction and calibration methodologies, including removal mechanisms of instrumental artifacts and for instrument design. A significant application of {\ml} has been as a cross-instrument calibrator and for development of {\it virtual instruments} in domains with either lost coverage or no coverage. However, such translation and restoration operators are scientifically useful only when validated in physics space, not just in image space. Better visual perceptual quality is insufficient if flux conservation, polarity consistency, thermodynamic plausibility, or downstream forecast skill are degraded. The field urgently needs harmonized validation frameworks and reproducible benchmark datasets that allow fair model comparison across missions and epochs.
\end{itemize}

These points lead to a practical agenda for the next phase of solar ML:
\begin{itemize}
    \item Move from point estimates to probabilistic operators with calibrated posteriors and explicit failure modes: Given that the forward problem in solar physics is always some form of projection operation, probabilistic estimate of the full distribution of the estimated parameters is needed from future applications of {\ml}.
    \item Prioritize hybrid physics-ML formulations that encode symmetries, conservation constraints, and observation geometry: Explicit encoding of known physical constraints is imperative in {\ml} modeling of the Sun.
    \item Design models and datasets for non-stationarity: The instruments performing measurements show a general degradation in time. Even if the instruments would be perfect, the solar cycle is not the same every time. This renders the system potentially non-ergodic, and requires adaptation of model to changing times~\citep{2024AGUFMSM11C..06J} 
    \item Treat interpretability as a physical requirement: The learned features must be testable against known mechanisms and hypothesis, and not just visually plausible~\citep{2020SpWea..1802478U}.
    \item  Close the research-to-operations loop with transparent reporting standards, robust all-clear evaluation, and stakeholder-specific metrics.
\end{itemize}
The goal is no longer to show that ML can emulate a task, but to establish which learned operators remain physically valid under distribution shift, sparse supervision, and incomplete observability. This brings us to multi-modal foundation models, which are trained on large sets of multimodal data, and can generate powerful representations of data by cross-correlating across space, time and wavelength. Pioneering efforts such as Solaris~\citep{majid2024solaris}, and the recently developed Surya foundation model~\citep{surya} leverage massive, multi-instrument archives to learn unified, physics-aware representations. Such models are then very quickly fine-tuned for any other downstream tasks, and demonstrate that this transition toward fully coupled, data-driven modeling of the Sun-to-Earth system is already successfully underway. Their their long-term value will depend on whether they improve causal understanding, uncertainty-aware forecasting, and coordinated Sun-to-Earth decision making. However, the end goal would be to have a unified embedding space, where different models would interact with each-other, working with different data modalities, and moving to an embedding space that would be a very strong approximation of the state vector that quantifies every aspect of the Sun.


\begin{acknowledgement}
 VU would like to acknowledge NASA for support under award number 80NSSC25K7956 in support for writing this chapter. The authors acknowledge the use of AI tools for data collection and compilation. The agentic AI pipeline is made available on Github. 
\end{acknowledgement}





\bibliographystyle{custom_apalike}

\bibliography{sections/1_dynamical_system,%
sections/2_taxonomy,%
sections/3_interior_flows,%
sections/3_interior_magnetic,%
sections/3_interior_dynamo,%
sections/4_atmosphere,%
sections/4_atmosphere_largescale,%
sections/5_phenomena,%
sections/6_observations_cluster1,%
sections/6_observations_cluster2,%
sections/6_observations_cluster3,%
sections/7_conclusion%
}



\end{document}